\documentclass[namedreferences]{solarphysics}

\usepackage[hyperref,optionalrh,solaromanenum]{spr-sola-addons} 
\usepackage[graphicx]{realboxes}
\usepackage{color}           
\usepackage{breakurl}        
\usepackage{lscape}
\usepackage{longtable}

\newcommand{\apj}{    {\it Astrophys. J.}}

\chardef\us=`\_

\begin{document}

\begin{article}
\begin{opening}

\title{A Catalog of Bipolar Active Regions Violating the Hale Polarity 
Law, 1989\,--\,2018}

\author[addressref=aff1,corref,email={anastasiya.v.zhukova@gmail.com}]
{\inits{A.V.}\fnm{Anastasiya}~\lnm{Zhukova}\orcid{0000-0003-4435-6706}}
\author[addressref=aff2,email={hlystova@iszf.irk.ru}]
{\inits{A.I.}\fnm{Anna}~\lnm{Khlystova}\orcid{0000-0001-9907-5392}}
\author[addressref=aff1,email={vabramenko@gmail.com}]
{\inits{V.I.}\fnm{Valentina}~\lnm{Abramenko}\orcid{0000-0001-6466-4226}}
\author[addressref={aff3,aff4,aff5},email={sokoloff.dd@gmail.com}]
{\inits{D.D.}\fnm{Dmitry}~\lnm{Sokoloff}}
\address[id=aff1]{Crimean Astrophysical Observatory of Russian Academy 
of Sciences, Nauchny 298409, Bakhchisaray, Republic of Crimea}
\address[id=aff2]{Institute of Solar-Terrestrial Physics of Siberian Branch 
of Russian Academy of Sciences, Lermontov st., 126a, Irkutsk 664033, Russia}
\address[id=aff3]{Department of Physics, Lomonosov Moscow State University, 
Faculty of Physics, Leninskie Gory, 1-2, Moscow 119992, Russia}
\address[id=aff4]{Moscow Center of Fundamental and Applied Mathematics, 
Moscow 119991, Russia}
\address[id=aff5]{IZMIRAN, Kaluzhskoe shosse, 4, Troitsk, Moscow,  142191, 
Russia}

\runningauthor{A.V.~Zhukova et al.}
\runningtitle{\textit{Solar Physics} A catalog of anti-Hale ARs}

\begin{abstract}
There is no list of bipolar active regions (ARs) with reverse polarity 
(anti-Hale regions), although statistical investigations 
of such ARs (bearing the imprint of deep subphotospheric processes) are 
important for understanding solar-cycle mechanisms. We studied 8606 ARs 
from 1 January 1989 to 31 December 2018 to detect anti-Hale regions and 
to compile a catalog. The {\it Solar and Heliospheric Observatory} (SOHO) and 
the {\it Solar Dynamics Observatory} (SDO) data, as well as the Debrecen 
Photoheliographic Data, the Mount Wilson Observatory catalog and drawings, 
and the USAF/NOAA Solar Region Summary were used. Complex, ambiguous cases 
related to anti-Hale region identification were analyzed. Two basic and 
four additional criteria to identify an AR as an anti-Hale region 
were formulated. The basic criteria assume that: i) dominating features 
of an AR have to form a bipole of reverse polarity with sunspots/pores 
of both polarities being present; ii) magnetic connections between 
the opposite polarities has to be observed. A catalog of anti-Hale 
regions (275 ARs) is compiled. The catalog contains: NOAA number, 
date of the greatest total area of sunspots, coordinates, 
and corrected sunspot area for this date. The tilt and the most complex 
achieved Mount Wilson magnetic class are also provided. The percentage 
of anti-Hale groups meeting the proposed criteria is $\approx$3.0\,\% 
from all studied ARs, which 
is close to early estimations by authors who had examined each AR 
individually: $\approx$2.4\,\% by Hale and Nicholson 
(\apj{} \textbf{62}, 270, \citeyear{Hale25}) and $\approx$3.1\,\% 
by Richardson (\apj{} \textbf{107}, 78, \citeyear{Richardson48}). 
The enchancement of the anti-Hale percentage in later research might be 
related to: i) increasing sensitivity of instruments (considering smaller 
and smaller bipoles); ii) the ambiguities in the anti-Hale region 
identification. 
The catalog is available as the Electronic Supplementary Material and
at the CrAO website (sun.crao.ru/databases/catalog-anti-hale/).

\end{abstract}
\end{opening}

\section{Introduction}

\begin{figure}     
\centerline{\hspace*{0.015\textwidth}
\includegraphics[width=0.25\textwidth,clip=]{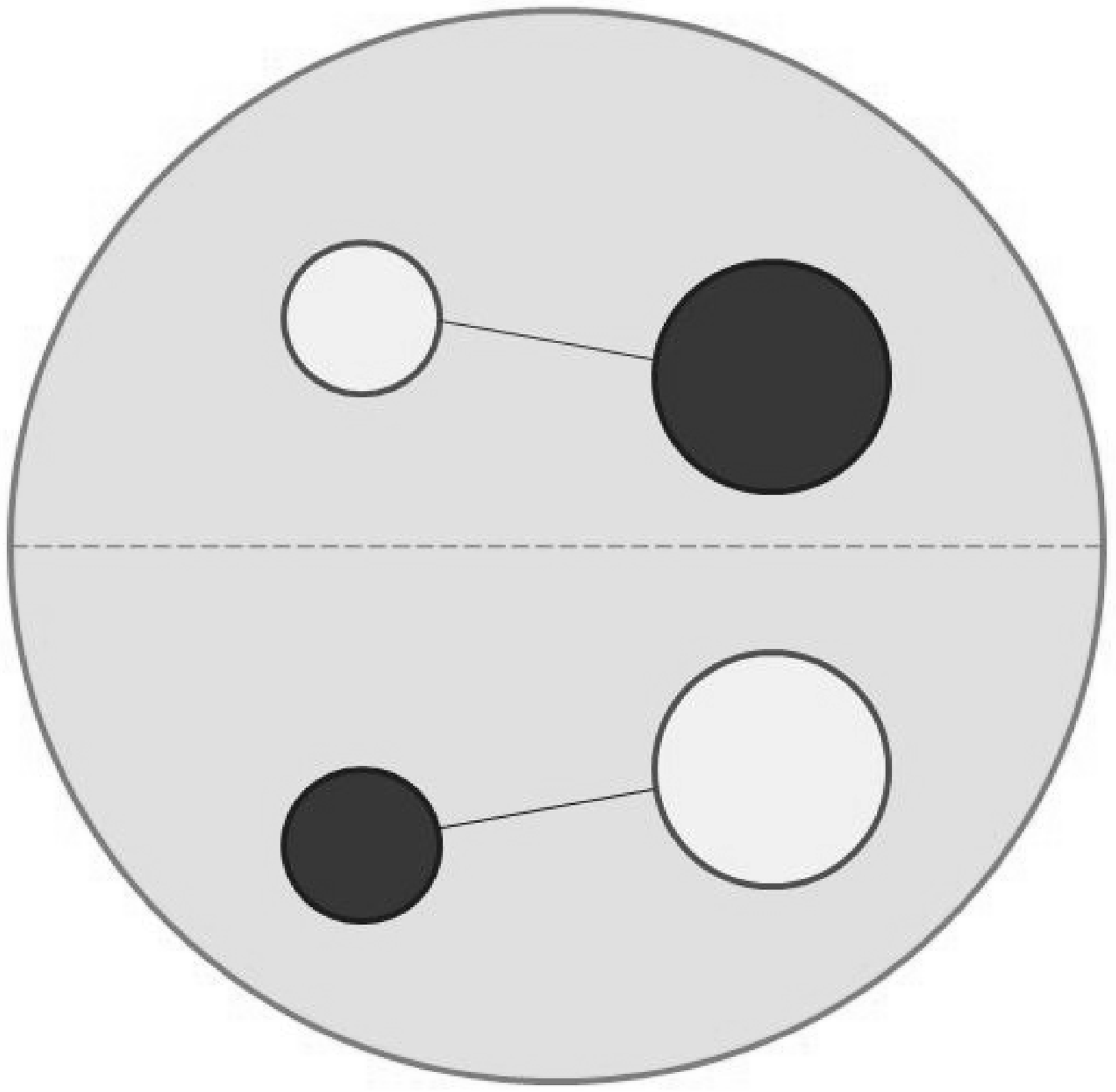}
\hspace*{0.015\textwidth}
\includegraphics[width=0.245\textwidth,clip=]{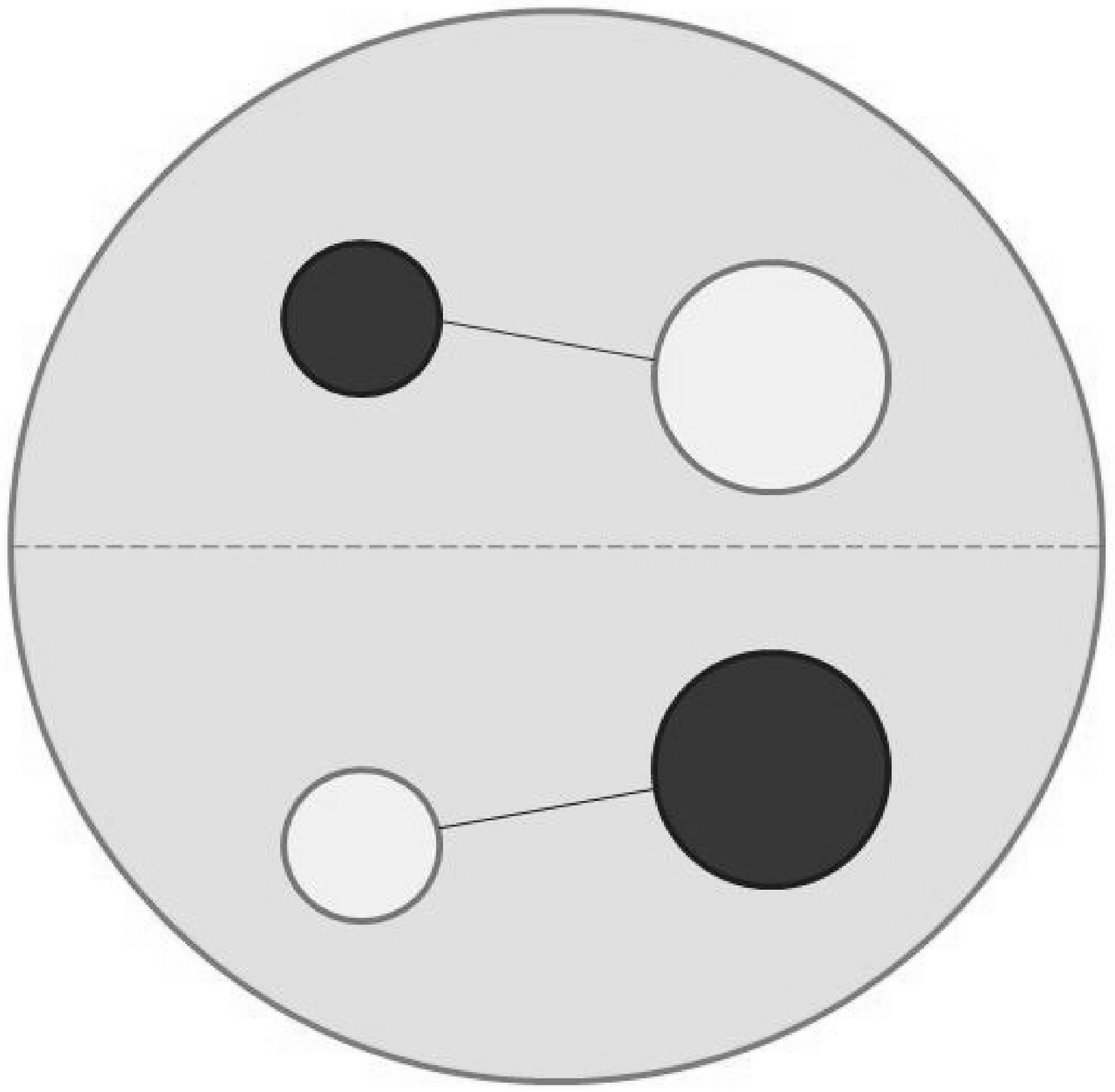}
}

\vspace{-0.25\textwidth}   
\centerline{\Large \bf     
\hspace{0.22 \textwidth}  \color{black}{(a)}
\hspace{0.175\textwidth}  \color{black}{(b)}
\hfill}
\vspace{0.2\textwidth}    
                         
\caption{Regular (with the leading sunspot obeying Hale's 
polarity law) bipolar ARs in even (a) and odd (b) cycles. Sunspots 
with negative (positive) polarity are shown as black (white).
The polarity of sunspots of anti-Hale regions is opposite.}
\label{Fig_1}
\end{figure}

The majority of active regions (ARs) obey the empirical rules for sunspot 
groups \citep{Hale19}: Hale's polarity law, Joy's law, etc. (regular ARs). 
The pioneering models of the magnetic cycle (\citealp{Parker55}; 
\citealp{Babcock61}; \citealp{Leighton64}), as well as the global 
mean-field dynamo theory (see, e.g., \citealp{Moffatt78}; 
\citealp{Krause80}), explain well the existance of the regular ARs. 
However, ARs 
violating these empirical rules also occur. According to Hale's law during 
an even (odd) sunspot cycle, the leading sunspot of a bipolar AR has negative 
(positive) polarity in the northern hemisphere and the situation is reversed  
in the southern hemisphere. A sketch of regular ARs of even (odd) cycles 
is presented as Figure~\ref{Fig_1}. In this article, we deal with bipole sunspot 
groups with the reverse polarity (anti-Hale regions). 
We do not consider here other types of irregularities. 

Anti-Hale ARs bear the imprint of deep subphotospheric processes. So,  
statistical studies of these ARs may shed light on our understanding 
of solar magnetic-field evolution and solar-cycle mechanisms. 
For instance, \citet{Sokoloff15} found that a contribution of small-scale 
fluctuation dynamo in the interplay with the global dynamo is more 
pronounced at solar-cycle minimum. The authors came to this 
conclusion based on their probabilistic model and the revealed enhancement 
of the relative number of anti-Hale regions during solar minimum. 
Note that the presence of magnetic-field fluctuations 
on smaller (as compared to the mean-field) scales was earlier predicted 
by both the turbulent dynamo theory (e.g. \citealp{Kraichnan67}; 
\citealp{Kazantsev68}; \citealp{Zeldovich90}; \citealp{Brandenburg05}) and 
direct numerical simulations (e.g. \citealp{Blackman02}; 
\citealp{Radler03}; \citealp{Sur08}; \citealp{Brandenburg12}). 
The decay of anti-Hale ARs can explain the weakening of the polar magnetic 
field: the transport of remnants of the following parts of anti-Hale 
regions to the poles causes the decrease of the polar field 
\citep{Mordvinov19}. Numerical simulations by \citet{Hazra17} show that 
the high-latitude anti-Hale ARs in the middle phase of a cycle must 
contribute to weakening of the polar field more than the rest of anti-Hale 
groups. \citet{Jiang15} also found that the magnetic-flux advection through 
the Equator due to decay of anti-Hale ARs causes the weakening of
the process of the polar-field recovery. Determination of the magnetic twist 
and writhe for several ARs attributed as anti-Hale regions allowed 
\citet{LopezFuentes00,LopezFuentes03} to estimate the degree of influence  
of various mechanisms (kink-instability, effect of the Coriolis force, 
etc.) on the deformation of flux-tubes. Thus, the observational data 
on the statistics and properties of anti-Hale groups is in high demand 
in the solar physics community.

Unfortunately, there is no reliable database of anti-Hale 
regions. Although many catalogs provide information on the polarity of leading 
and following parts of ARs and use this information (for example, 
to measure the tilt of the sunspot-group axis), the catalogs do not provide 
these data explicitly. 
The Mount Wilson Observatory (MWO) catalog had been a main data source 
on the anti-Hale ARs for a long time. The catalog had been replenished 
irregularly since 1920. The data are available digitally since 1962.
From 1989 to 2004 the ARs of reverse polarity are marked with the sign 
``+''. According to many researchers who examined the catalog, some ARs 
in the catalog are marked with this sign erroneously. Thus, 
\citet{Hale25} viewed a lot of AR drawings to verify the magnetic class 
detected by MWO operators. \citet{Richardson48} reported checking 
each AR of reverse polarity from the MWO catalog based on the initial 
records. Errors in anti-Hale region identification in the MWO catalog 
were also revealed by \citet{Sokoloff15}. 

The published estimation of the relative number of anti-Hale regions varies 
within a rather wide range: from a few percent to approximately 
(8\,--\,9)\%. 
\citet{Hale25} found that $<$6 (2.4 on average) of all ARs are 
the reverse-polarity groups with simple bipolar magnetic configuration. 
Later the following results for this parameter were reported: 3.1 
\citep{Richardson48}, $<$5 \citep{Smith68}, 1.8 \citep{Vitinsky86}, 
4 \citep{Wang89}; 4.9 (\citealp{Khlystova09}, \citealp{Sokoloff10}), 
4 \citep{Stenflo12}, $\approx$8 (\citealp{Li12}; \citealp{McClintock14}; 
\citealp{Li18}). Note that only part of these studies were performed 
on the basis of the MWO catalog. Many authors used their own databases.
\citet{McClintock14} suggested that the inhomogeneity of the data used
might be the reason for such a discrepancy in the results of various 
authors. The size of included ARs is one of the main reasons 
for discrepancy, because small sunspot groups are prone to violate 
the Hale's polarity law more frequently (\citealp{Richardson48}; 
\citealp{Howard89}; \citealp{Stenflo12}). Only $\approx$60\,\% of ephemeral 
regions obey this law \citep{Hagenaar01}. Moreover, the smallest 
quiet-Sun bipoles are randomly oriented \citep{Tlatov10}. 

Errors in the identification of anti-Hale regions also occure. 
For example, there is an obvious difference in numbers and coordinates 
of anti-Hale ARs on the time--latitude diagrams reported 
by \citet{Richardson48} and by \citet{Vitinsky86} in spite of the fact 
that the authors used the same database. The known reasons for anti-Hale 
region misidentification are:
i) erroneous detection of ARs boundaries \citep{Sokoloff15}; 
ii) the wrong distribution of ARs between cycles during the transition 
from one solar cycle to anoter; iii) the misalignment between the magnetic 
and the heliographic Equators. The two latter reasons are discussed 
by \citet{McClintock14}. 

When compiling a database for further identification of anti-Hale 
regions, the type of the source data (magnetograms, drawings, etc.) 
is also important. 
Thus, \citet{Stenflo12} reported about the objective difficulties 
in automatic detection of individual sunspot groups from full-disk 
magnetograms and extraction of single ARs. Besides, inaccuracies related 
to detection of weighted centers of polarities may also produce errors. 
\citet{Wang15} revealed that white-light measurements 
generally underestimate the tilts (axial inclinations) of ARs, even 
when the leading- and the following-polarity sectors are identified 
correctly. 

To summarize, studies of anti-Hale regions are essential for understanding 
the solar dynamo. So far, there is no reliable list of ARs
with reverse polarity. We undertake a compilation of such a database 
starting from 1989 and covering the last two and a half solar cycles. We paid 
special attention to the analysis of mistakes and ambiguities in identification 
of anti-Hale ARs.

\section{Data and Method}

\subsection{Data Sources}

To examine ARs visually, we used the full-solar-disk daily data: 
magnetograms, continuum and extreme ultraviolet (EUV) images, sketches 
and drawings, as well as different images of the individual ARs. 
We also used available catalogs containing the information on the leading 
sunspot polarity of bipolar ARs and other necessary data. 
We used the following sources:

-- the Michelson Doppler Imager (MDI: \citealp{Scherrer95}) and the Extreme 
Ultraviolet Imaging Telescope (EIT: \citealp{Moses97}) on board the Solar and 
Heliospheric Observatory (SOHO);

-- the Helioseismic and Magnetic Imager (HMI: \citealp{Scherrer12}) and 
the Atmospheric Imaging Assembly (AIA: \citealp{Lemen12}) onboard the Solar 
Dynamic Observatory (SDO); 

-- the Debrecen Photoheliographic Data (DPD)\\ 
\citep{Baranyi16} accessible from\\ 
fenyi.solarobs.csfk.mta.hu/DPD/;

-- the websites www.helioviewer.org (Helioviewer), www.solarmonitor.org 
(Solar Monitor); 

-- the MWO drawings - which are available in the digital archive at\\ 
ftp://howard.astro.ucla.edu/pub/obs/drawings/;

-- the MWO catalog accessible at the National Geophysical Data Center at\\
ftp://ftp.ngdc.noaa.gov/STP/SOLAR\_DATA/SUNSPOT\_REGIONS/\\
Mt\_Wilson/; 

-- the Crimean Astrophysical Observatory (CrAO) catalog\\ 
(\citealp{Abramenko18}; \citealp{Zhukova18}) at\\ 
sun.crao.ru/databases/catalog-mmc-ars;

-- the United States Air Force/National Oceanic and Atmospheric 
Administration Solar Region Summary (USAF/NOAA SRS) at\\ 
solarcyclescience.com/activeregions.html.

Note that at the biginning of our study we also used the Royal Observatory, 
Greenwich -- USAF/NOAA Sunspot Data provided by the National Aeronautics 
and Space Administration (NASA). But these data are available only until 
October 2016 (solarscience.msfc.nasa.gov/greenwch.shtml), 
and later we restricted ourselves to the USAF/NOAA SRS database.

\subsection{Ambiguities in Anti-Hale Region Identification}

\begin{figure}
\centerline{
\includegraphics[width=0.7\textwidth]{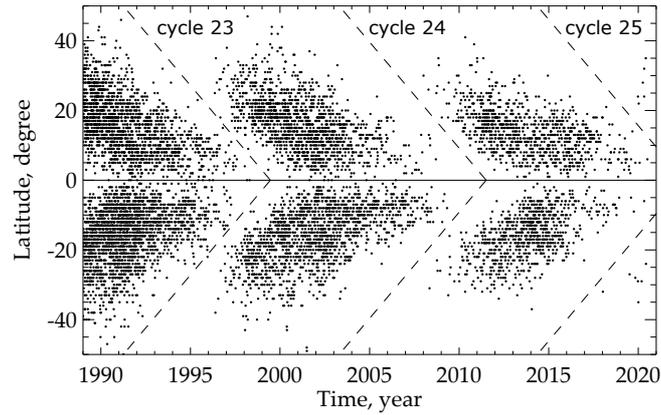}
}
\caption{Time--latitude diagram of sunspot groups
(by USAF/NOAA SRS data) with boundaries (dotted lines) between cycles 
plotted by \citet{McClintock14} technique.} 
\label{Fig_2}
\end{figure}

We encountered the following ambiguities in anti-Hale AR identification.

i) During solar-cycle minima the high-latitude sunspot groups (new-cycle 
precursors) might be erroneously accepted as reverse-polarity ARs. 
The periods of this ambiguity are 1996 and 2008 (the cycle minima). 
To determine what cycle a suspicious high-latitude group belongs to, 
we used a technique suggested by \citet{McClintock14}. The boundaries 
between cycles in North and South hemispheres were defined as: 
\begin{equation}  \label{Eq}
     60 \times ylat = \pm (xdate - xint),
   \end{equation}
where $ylat$ is latitude, $xdate$ is a function of a number of days
from 1 January 1974, and $xint$ (a point of inter section at the Equator)
was visually determined separately for the best fit for each cycle.
The time--latitude diagram based on the USAF/NOAA SRS sunspot-group
data is presented as Figure \ref{Fig_2}. Cycle boundaries for Solar Cycles 
22/23 and 23/24 are plotted from \citet{McClintock14} data. We also plotted 
here the fit of the boundary between Solar Cycles 24 and 25.  
This boundary might be slightly changed with Solar Cycle 25 evolution.

\begin{figure}
\centerline{
\includegraphics[width=0.325\textwidth]{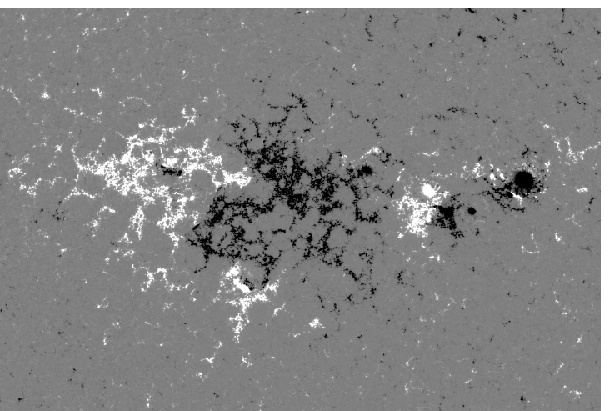}
\includegraphics[width=0.325\textwidth]{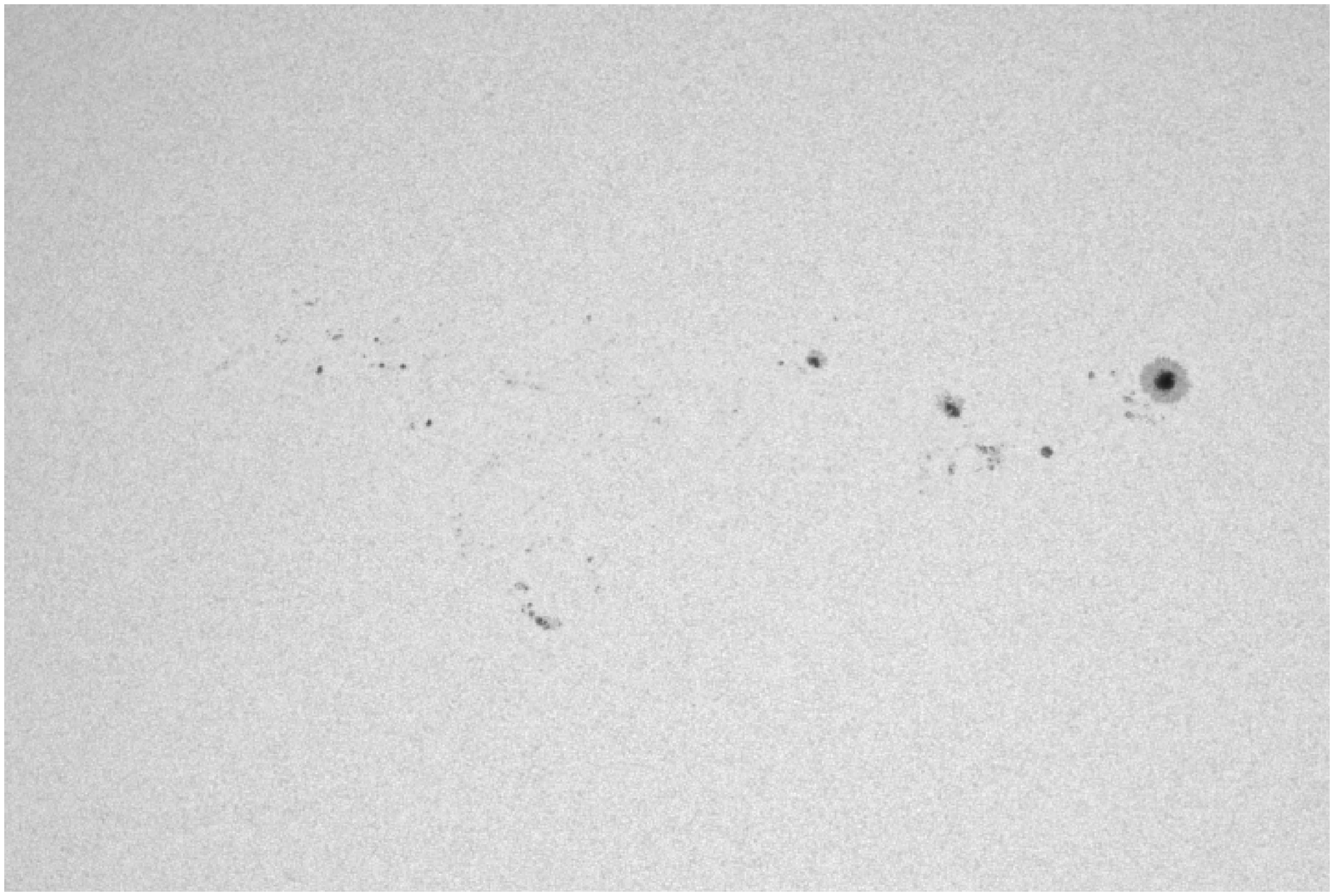}
\includegraphics[width=0.325\textwidth]{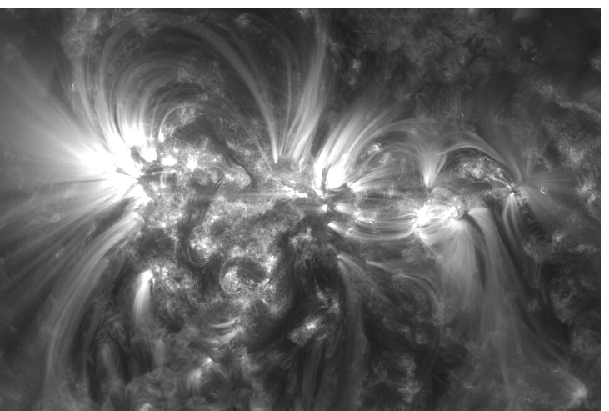}
}
\centerline{
\includegraphics[width=0.325\textwidth]{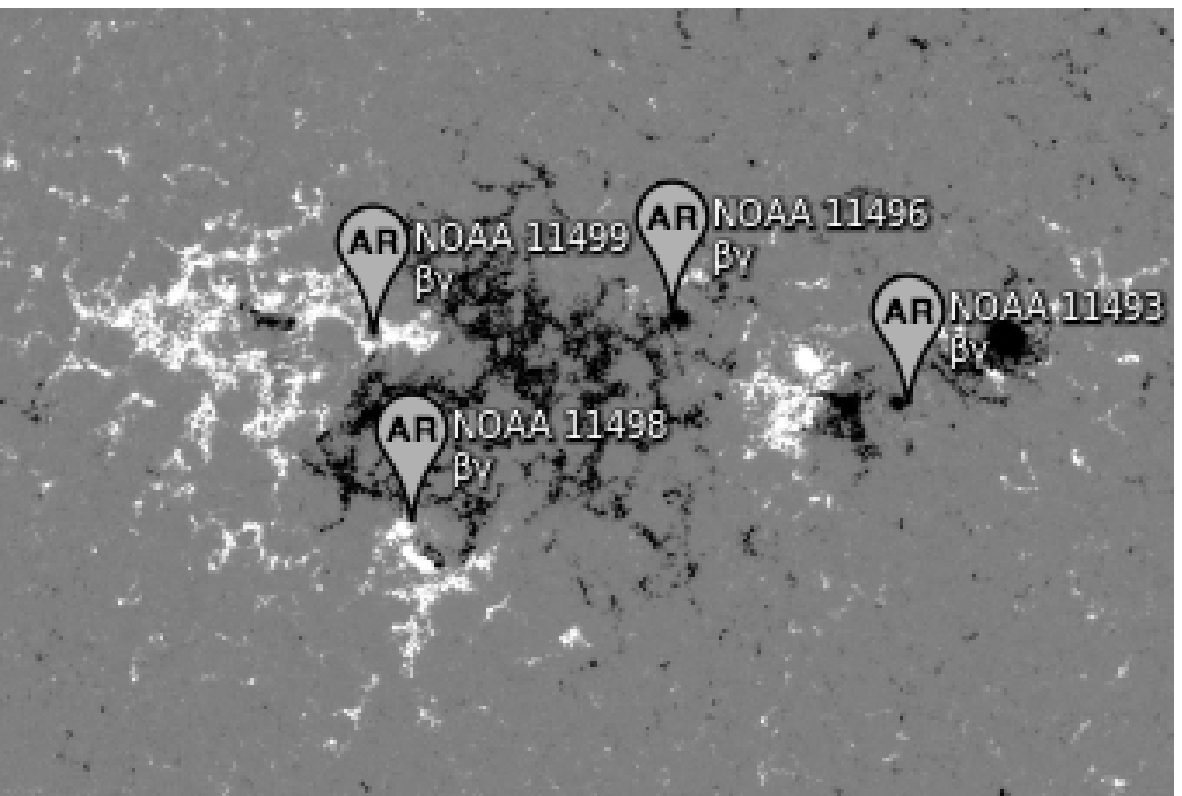}
\includegraphics[width=0.325\textwidth]{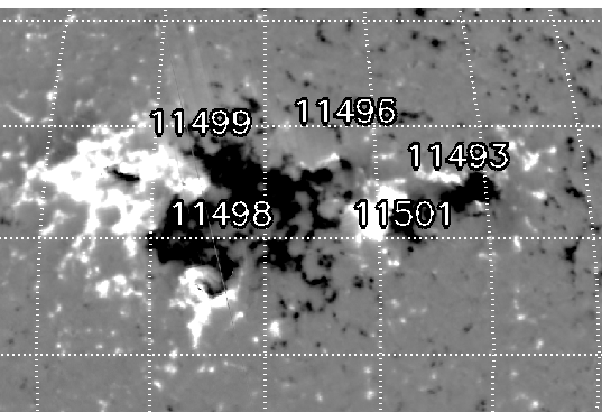}
\includegraphics[width=0.325\textwidth]{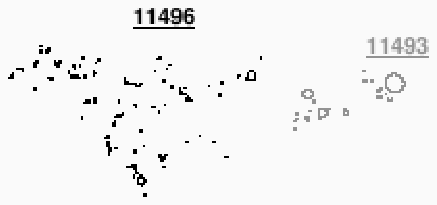}
}
\vspace{-0.29\textwidth}   
\centerline{\Large \bf     
\hspace{-0.008\textwidth}   \color{white}{(a)}
\hspace{0.245\textwidth}  \color{black}{(b)}
\hspace{0.248\textwidth}  \color{white}{(c)}
\hfill}

\vspace{0.192\textwidth}   
\centerline{\Large \bf     
\hspace{-0.008\textwidth}   \color{white}{(d)}
\hspace{0.245\textwidth}  \color{white}{(e)}
\hspace{0.248\textwidth}  \color{black}{(f)}
\hfill}

\caption{Composite activity complex observed on 06 June 2012 
in the North hemisphere: HMI magnetogram (a), continuum (b), and 
AIA Fe {\sc ix/x} 171~\AA\ line (c) images. Helioviewer (d), Solar 
Monitor (e), and DPD (f) interpretations of AR detection 
as an illustration of the difference in AR-separation algorithms. 
There are no anti-Hale regions in this block of ARs.}
\label{Fig_3}
\end{figure}

ii) Detection of individual sunspot groups in large composite 
activity complexes represents a challenge. \citet{Stenflo12} supposed 
that it does not seem feasible to make a program that can automatically 
and reliably identify {\it all} of the bipolar regions. Blocks and chains 
of ARs are typical for solar maxima. In that and some other cases 
sunspots/pores may be attributed with equal probability to several 
adjacent groups. Differences in segmentation of a magnetic complex 
by various tools is illustrated in Figure \ref{Fig_3}. We assumed that 
there are no anti-Hale sub-regions in this particular case.

\begin{figure}
\centerline{
\includegraphics[width=0.25\textwidth]{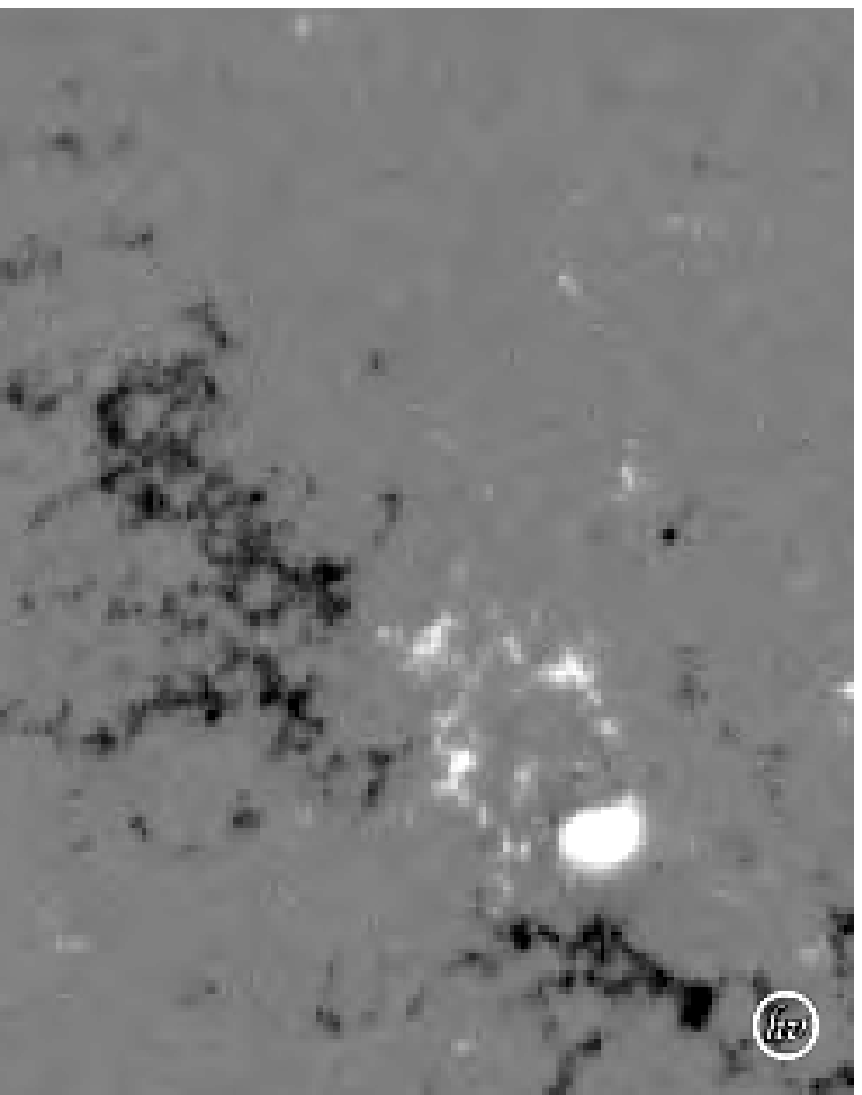}
\includegraphics[width=0.25\textwidth]{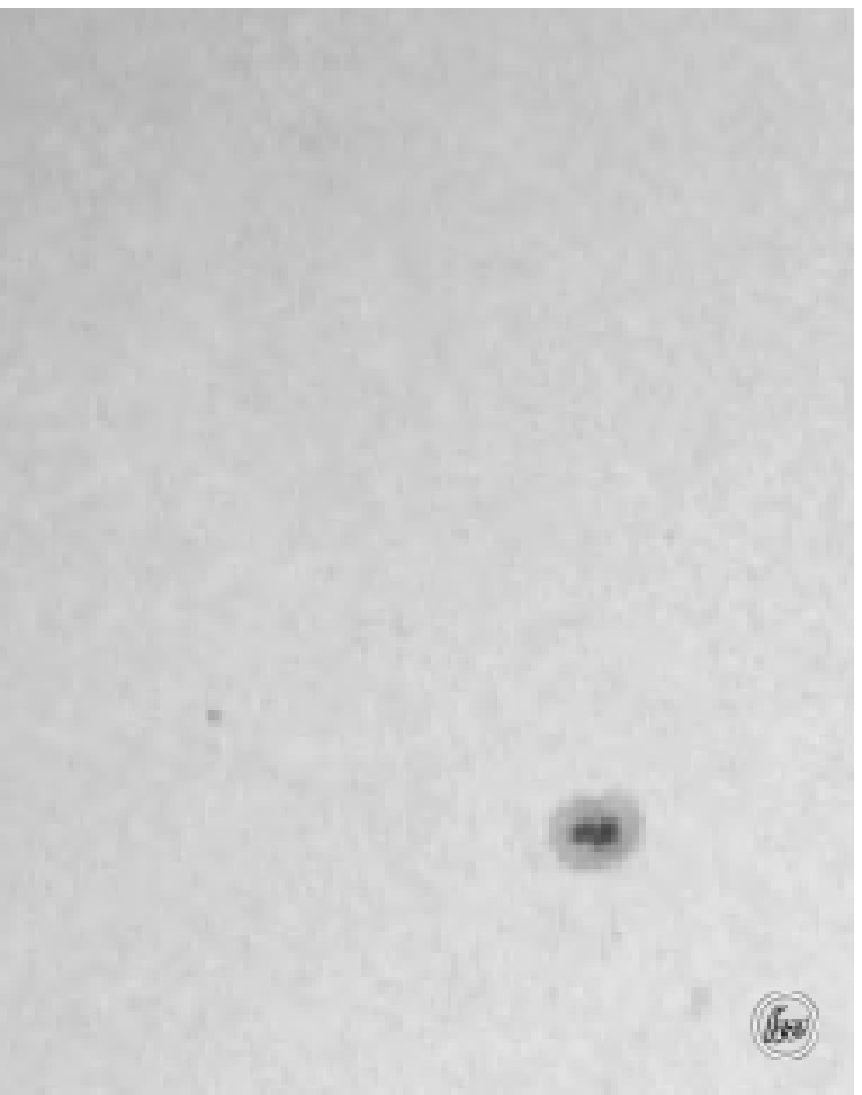}
\includegraphics[width=0.25\textwidth]{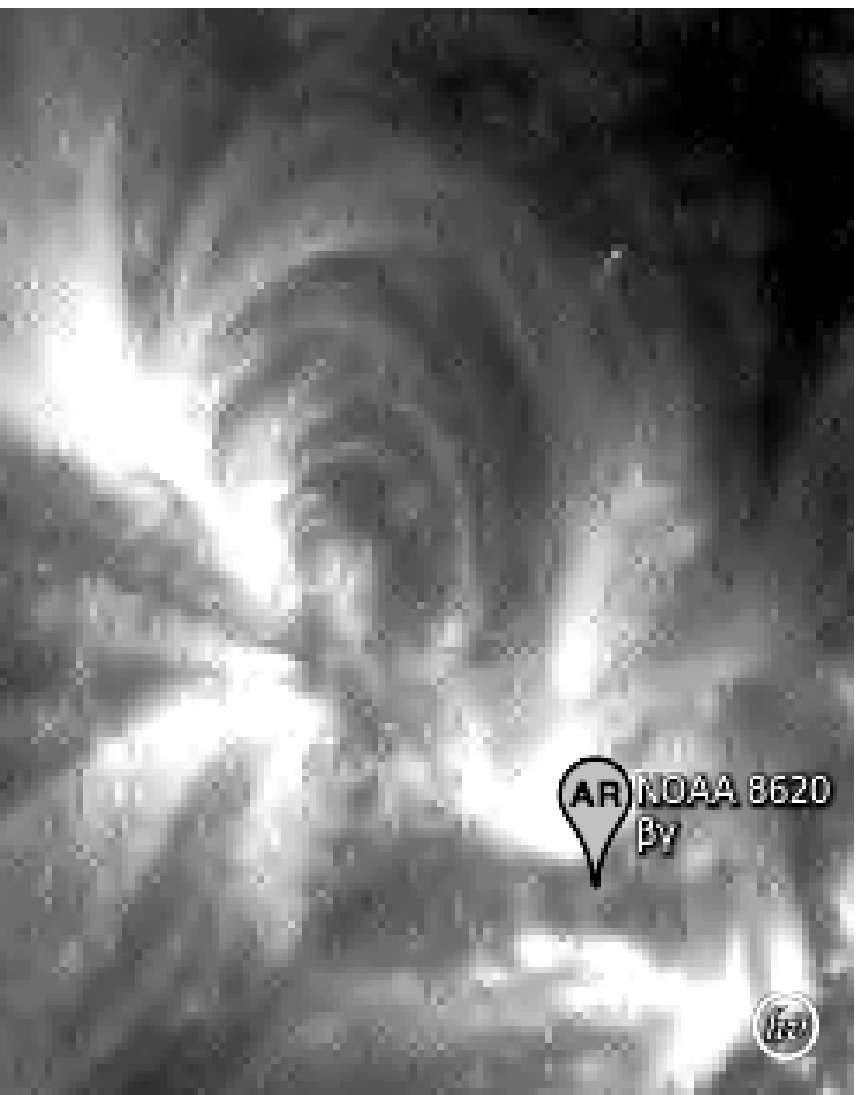}
}
\centerline{
\includegraphics[width=0.25\textwidth]{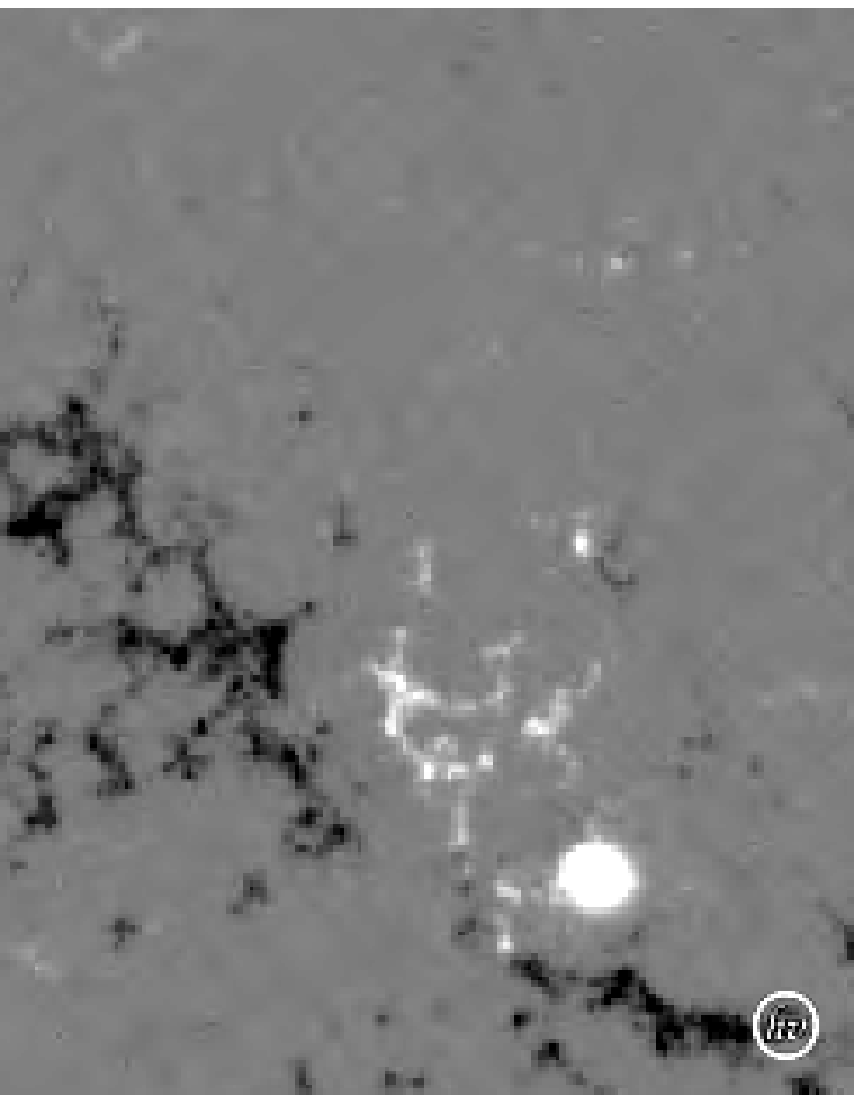}
\includegraphics[width=0.25\textwidth]{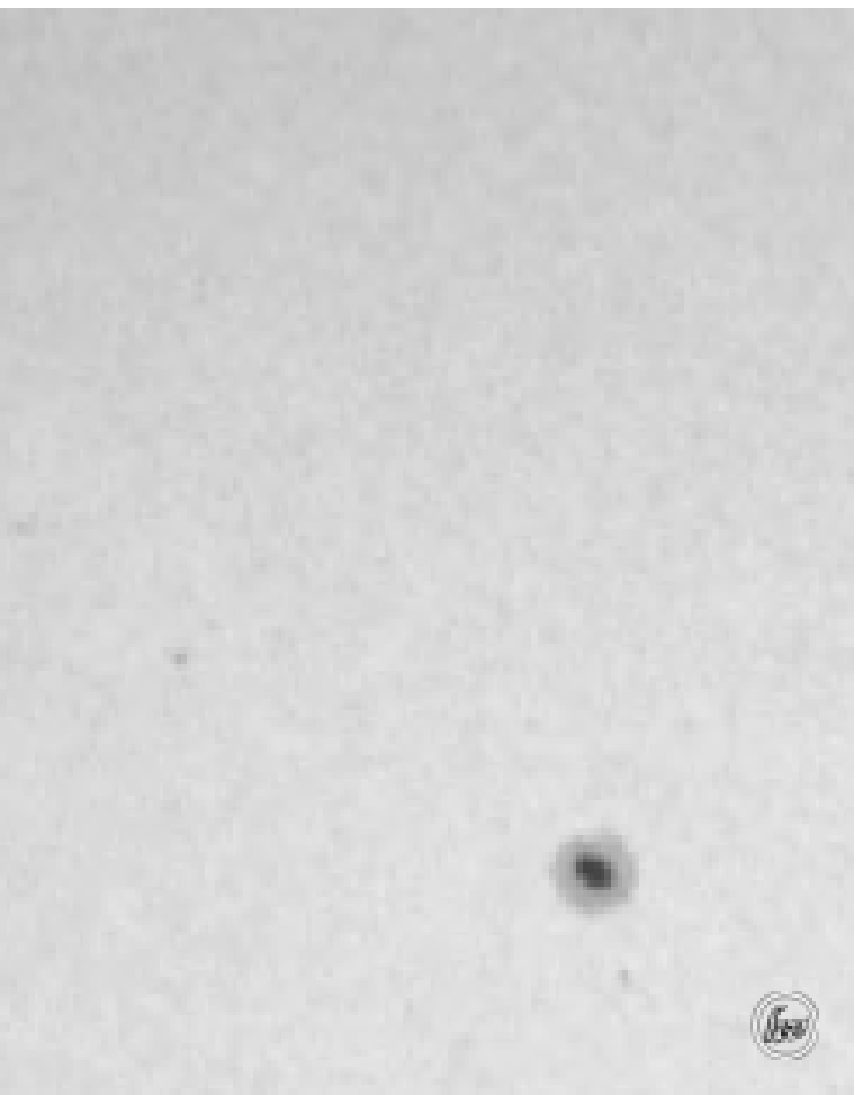}
\includegraphics[width=0.25\textwidth]{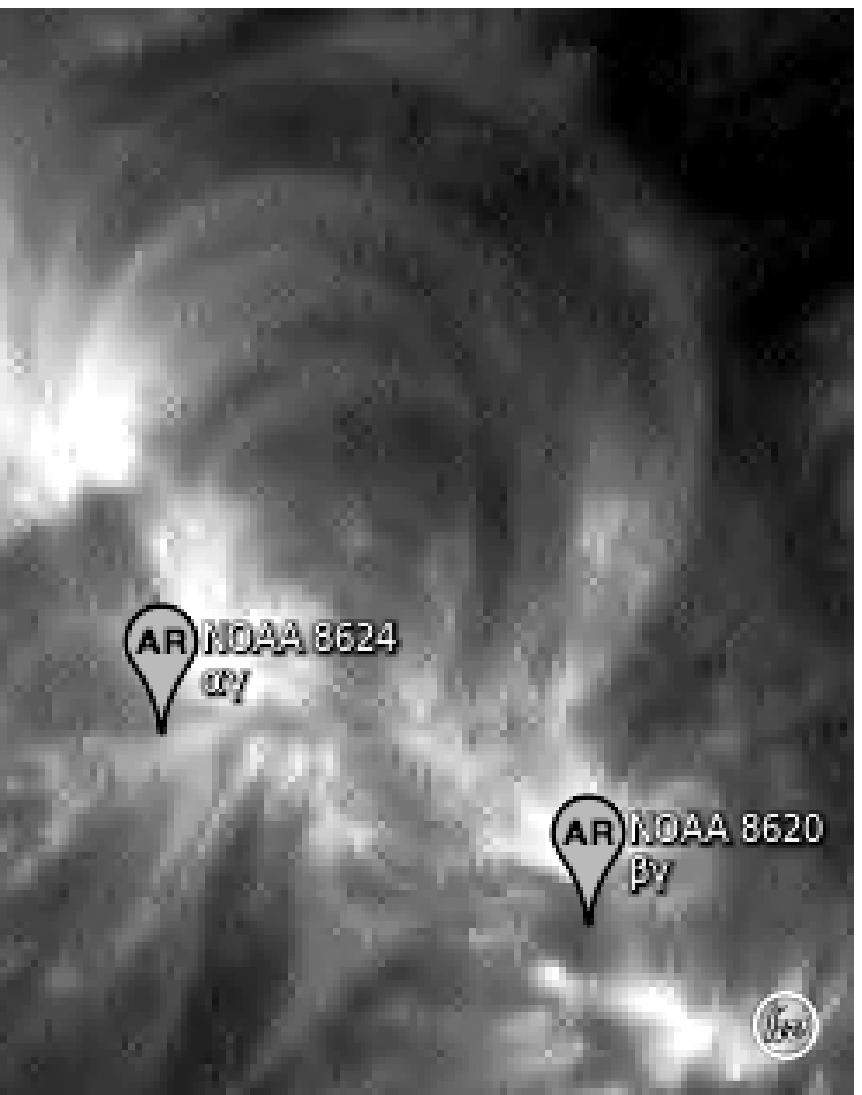}
}
\vspace{-0.38\textwidth}   
\centerline{\Large \bf     
\hspace{0.11\textwidth}   \color{white}{(a)}
\hspace{0.17\textwidth}  \color{black}{(b)}
\hspace{0.17\textwidth}  \color{white}{(c)}
\hfill}

\vspace{0.28\textwidth}   
\centerline{\Large \bf     
\hspace{0.11\textwidth}   \color{white}{(d)}
\hspace{0.17\textwidth}  \color{black}{(e)}
\hspace{0.17\textwidth}  \color{white}{(f)}
\hfill}                                  

\caption{Sunspot group in the North hemisphere assigned by Helioviewer 
as a single AR 8620 on 06 July 1999: HMI magnetogram (a), continuum (b), 
and AIA Fe {\sc ix/x} 171~\AA\ line (c) images; the same images of this sunspot
group on 07 July (d, e, f): two separate ARs (NOAA 8620 and 8624) are 
assigned. A part of the mafgnetic structure assigned as NOAA 8620 
on panel (f) is not a true anti-Hale AR because there is no magnetic 
connection between sunspots.}
\label{Fig_4}
\end{figure}

iii) Phenomena in ARs in the late decaying stage when the following spots 
of an AR have vanished or weakened:

a) Presence of opposite polarity small sunspots/pores to the West of an old 
leading sunspot. To illustrate, an example of 
a large decaying AR is shown in Figure~\ref{Fig_4}. Magnetic connections 
between a large leading spot of positive polarity and fragmented 
pores of negative polarity to the East are evident and steady (panels c, f). 
Such magnetic loops is the typical connection of the leading and 
the following parts of bipolar ARs. The absense of a magnetic 
connection between the large leading spot and a negative polarity pore 
to the South-West does not allow us to identify this couple of sunspots 
as a single AR with reverse polarity. Note, however, that this 
couple of sunspots is assigned by Helioviewer as 
an individual AR NOAA 8620 on 07, 09 July 1999 and further;
the following part of the initial sunspot group is extracted by Helioviewer 
as AR NOAA 8624. Such situation is typical for an AR with complex magnetic 
configuration (classified ``$\gamma$'' in the Mount Wilson class). 

\begin{figure}
\centerline{
\includegraphics[width=0.192\textwidth]{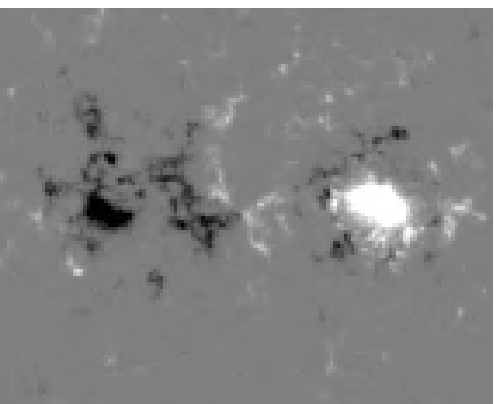}
\includegraphics[width=0.192\textwidth]{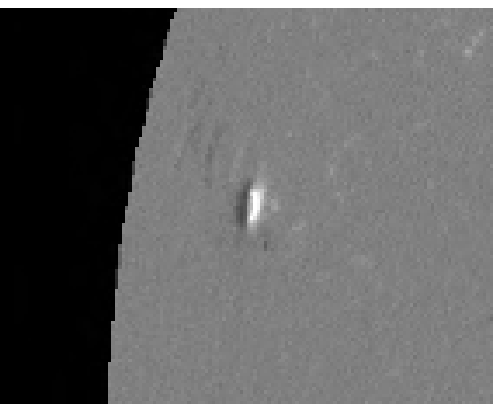}
\includegraphics[width=0.192\textwidth]{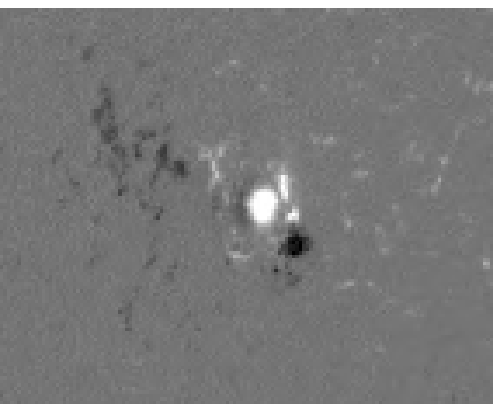}
\includegraphics[width=0.192\textwidth]{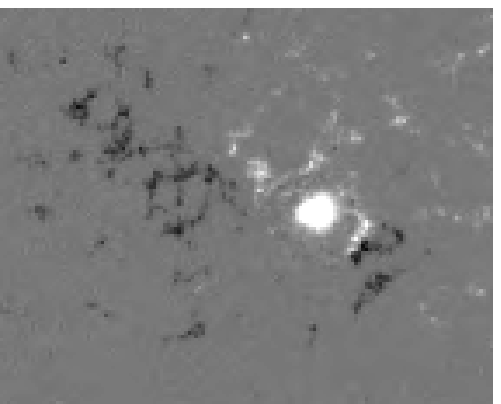}
\includegraphics[width=0.192\textwidth]{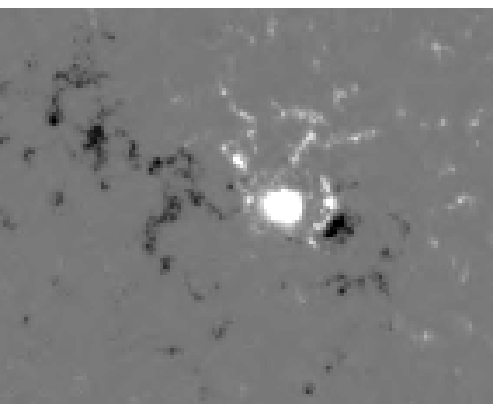}
}

\centerline{
\includegraphics[width=0.192\textwidth]{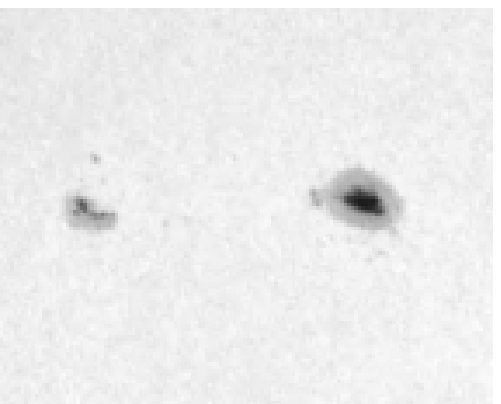}
\includegraphics[width=0.192\textwidth]{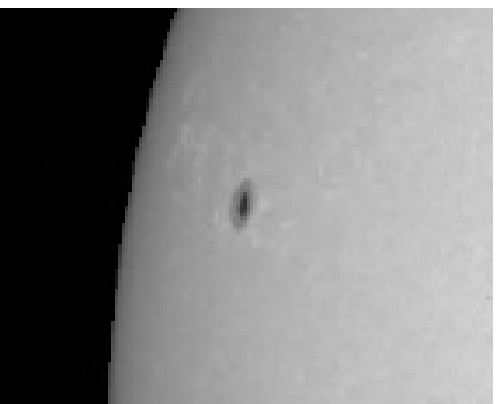}
\includegraphics[width=0.192\textwidth]{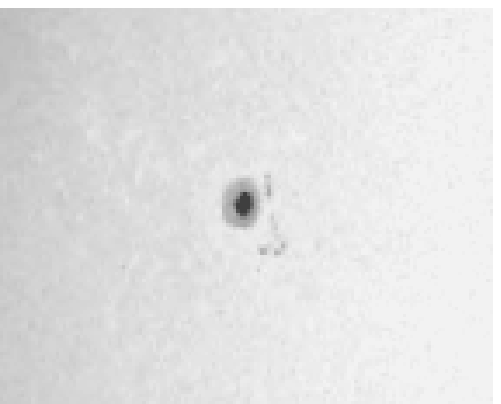}
\includegraphics[width=0.192\textwidth]{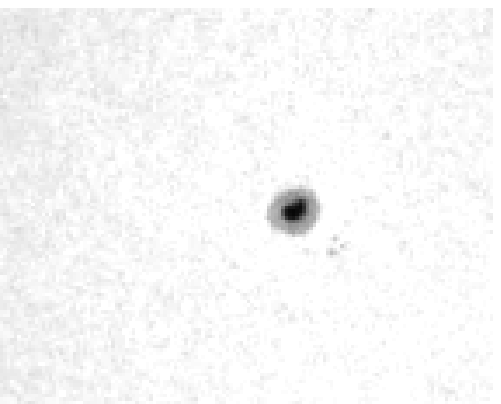}
\includegraphics[width=0.192\textwidth]{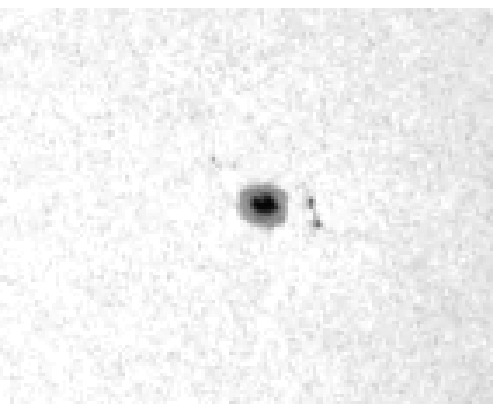}
}

\vspace{-0.215\textwidth}   
\centerline{\Large \bf     
\hspace{-0.01\textwidth}   \color{white}{(a)}
\hspace{0.115\textwidth}  \color{white}{(c)}
\hspace{0.115\textwidth}  \color{white}{(e)}
\hspace{0.11\textwidth}   \color{white}{(g)}
\hspace{0.11\textwidth}  \color{white}{(i)}
\hfill}

\vspace{0.122\textwidth}   
\centerline{\Large \bf     
\hspace{-0.01\textwidth}   \color{black}{(b)}
\hspace{0.115\textwidth}  \color{white}{(d)}
\hspace{0.115\textwidth}  \color{black}{(f)}
\hspace{0.11\textwidth}   \color{black}{(h)}
\hspace{0.11\textwidth}  \color{black}{(j)}
\hfill}

\caption{Recurrent AR NOAA 10634 (the North hemisphere) on 19 June 2004 (a,b) and 
its evolution as AR NOAA 10644 in the next rotation on 11 July 2004 (c,d), 
13 (e,f), 15 (g,h), 17 (i,j) illustrated by MDI magnetograms (upper panel) 
and continuum images (lower panel). Spots/pores in the following polarity 
of AR NOAA 10644 are absent; the decay of the long-living leading sunspot 
is accompanied by the formation of small spots/pores on the moat--cell 
boundaries. The combination of this with the old leading sunspot might be 
misidentified as an anti-Hale AR.}
\label{Fig_5}
\end{figure}

Moving magnetic features might also cause emerging of pores on moat--cell 
boundaries \citep{Ryutova18}. The combination of the old leading spot and\\ 
opposite-polarity small spot/pore to the West of it may be misidentified 
as a separate sunspot group violating Hale's polarity law 
(Figure~\ref{Fig_5}), especially by automatic algorithms.

b) Presence of opposite polarity pores all around the old sunspot. 
In cases when the flux (area) weighted center of scattered pores is located 
to the West of the old spot, the AR also might be erroneously identified 
as an anti-Hale region.

\begin{figure}
\centerline{
\includegraphics[width=0.192\textwidth]{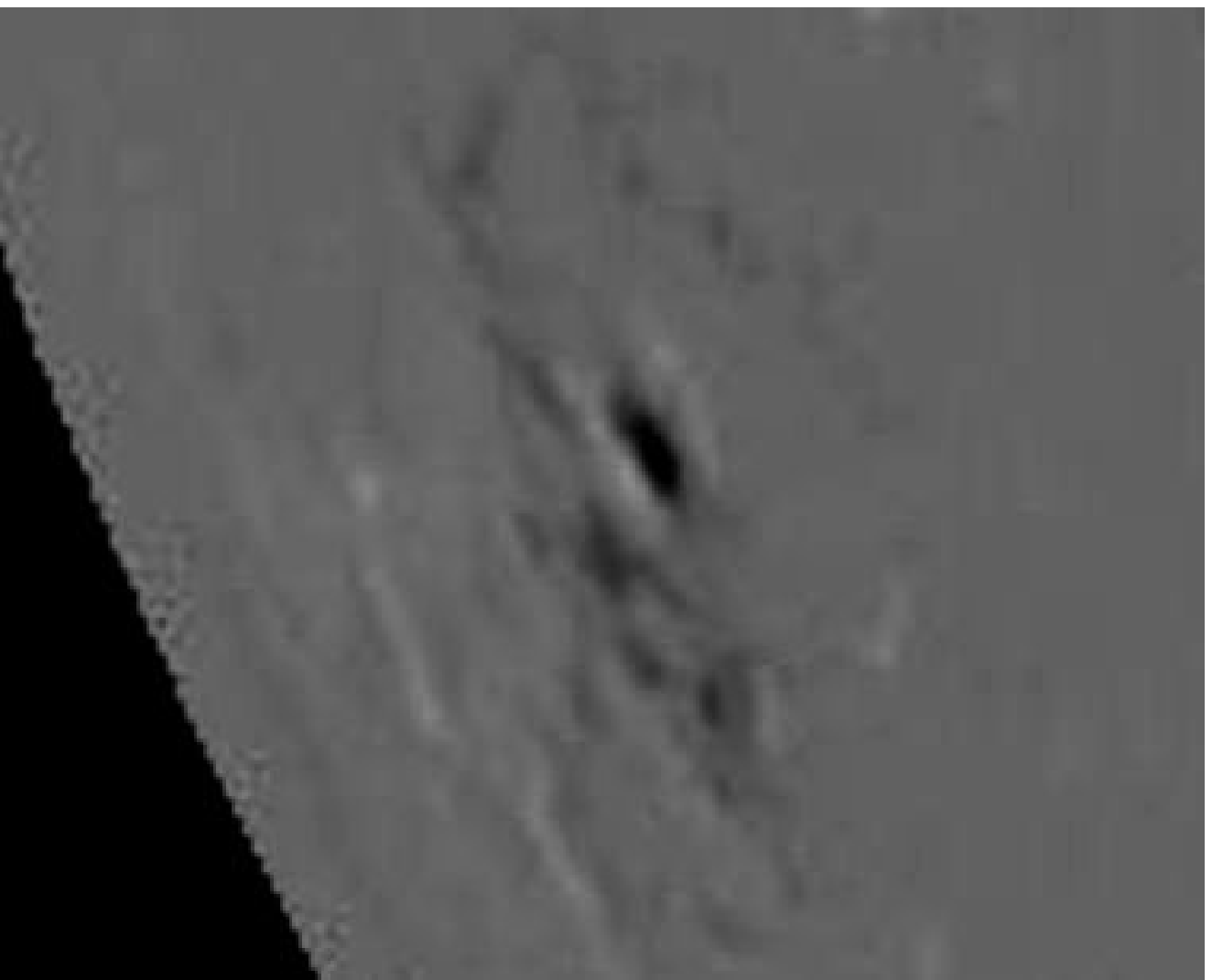}
\includegraphics[width=0.192\textwidth]{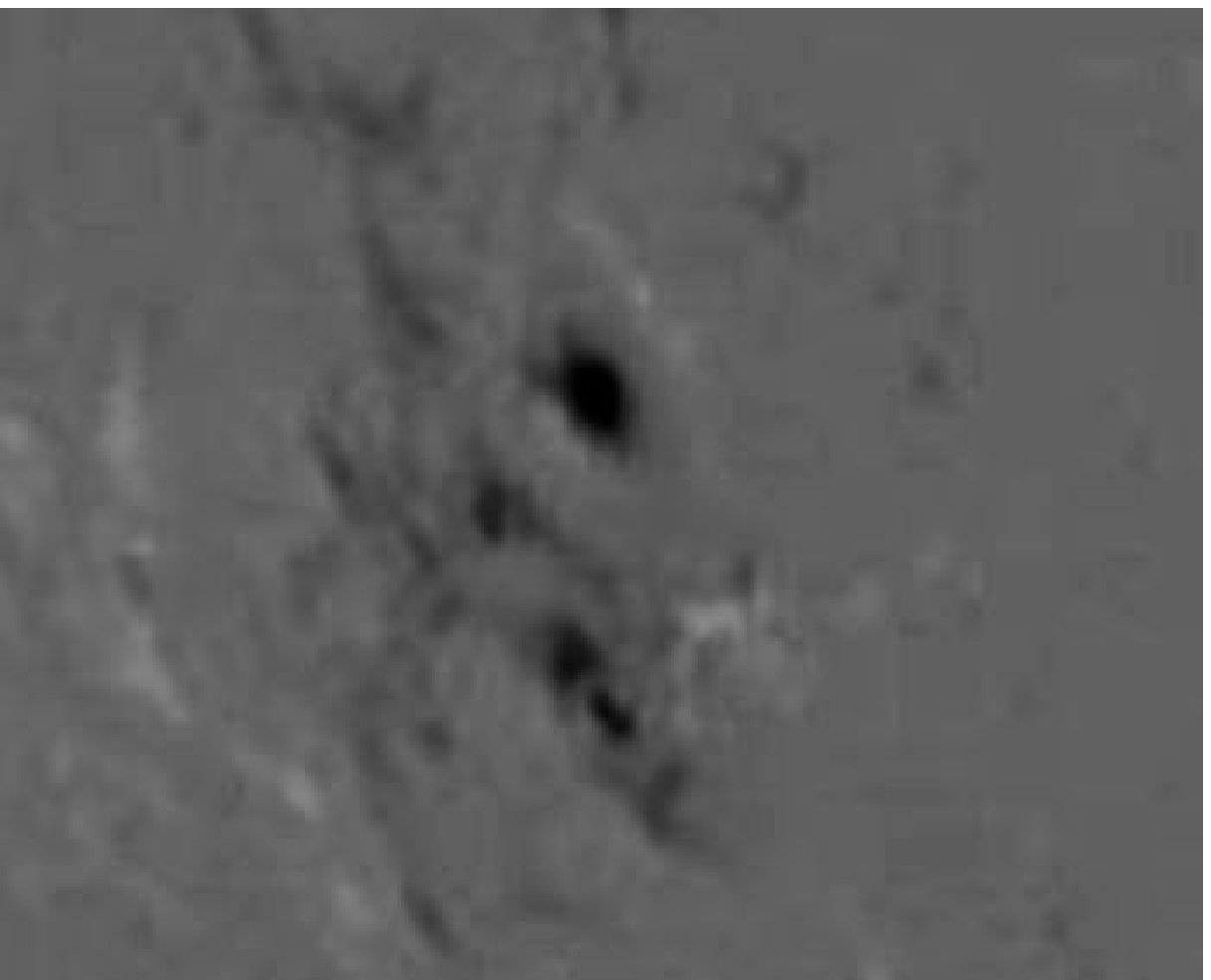}
\includegraphics[width=0.192\textwidth]{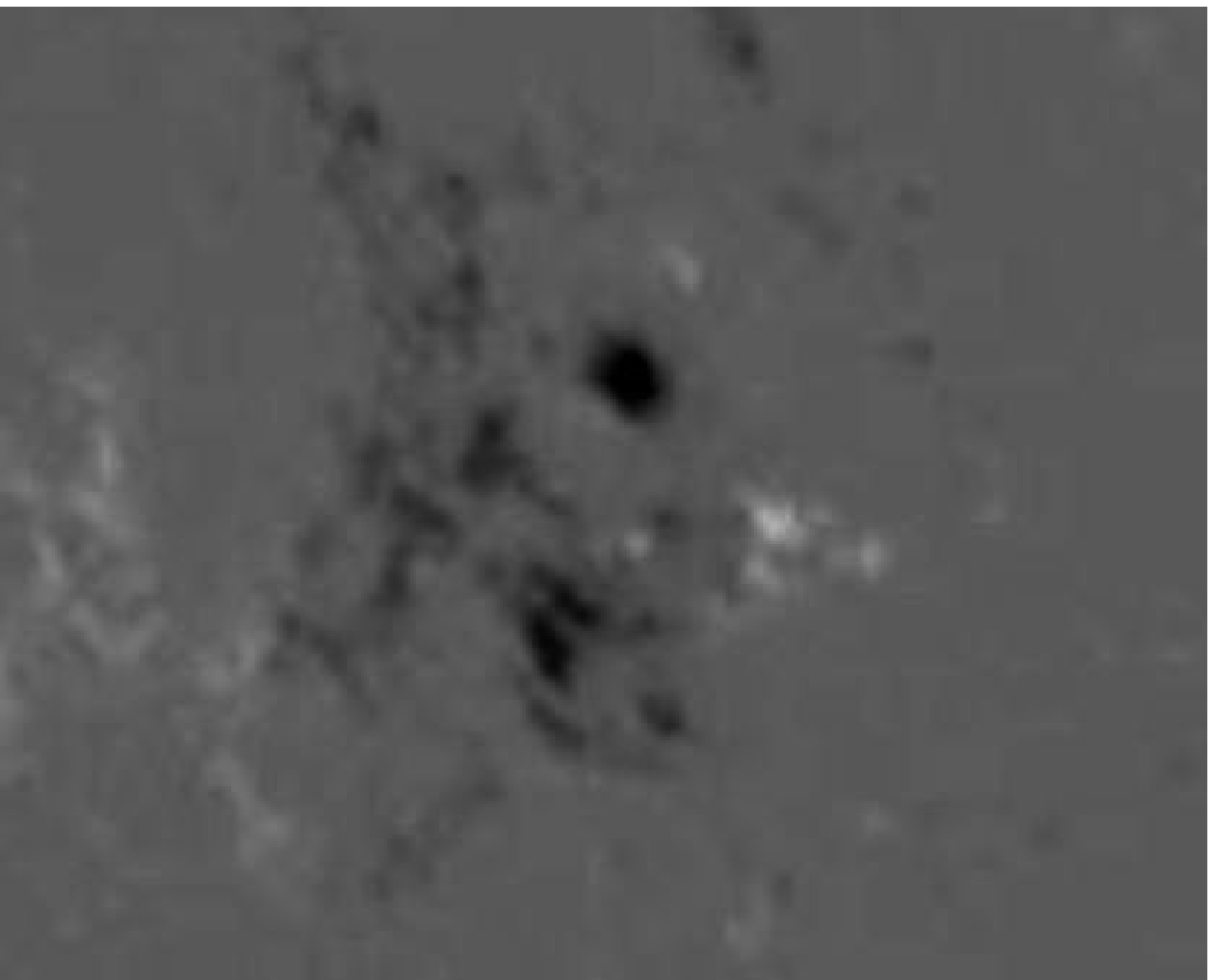}
\includegraphics[width=0.192\textwidth]{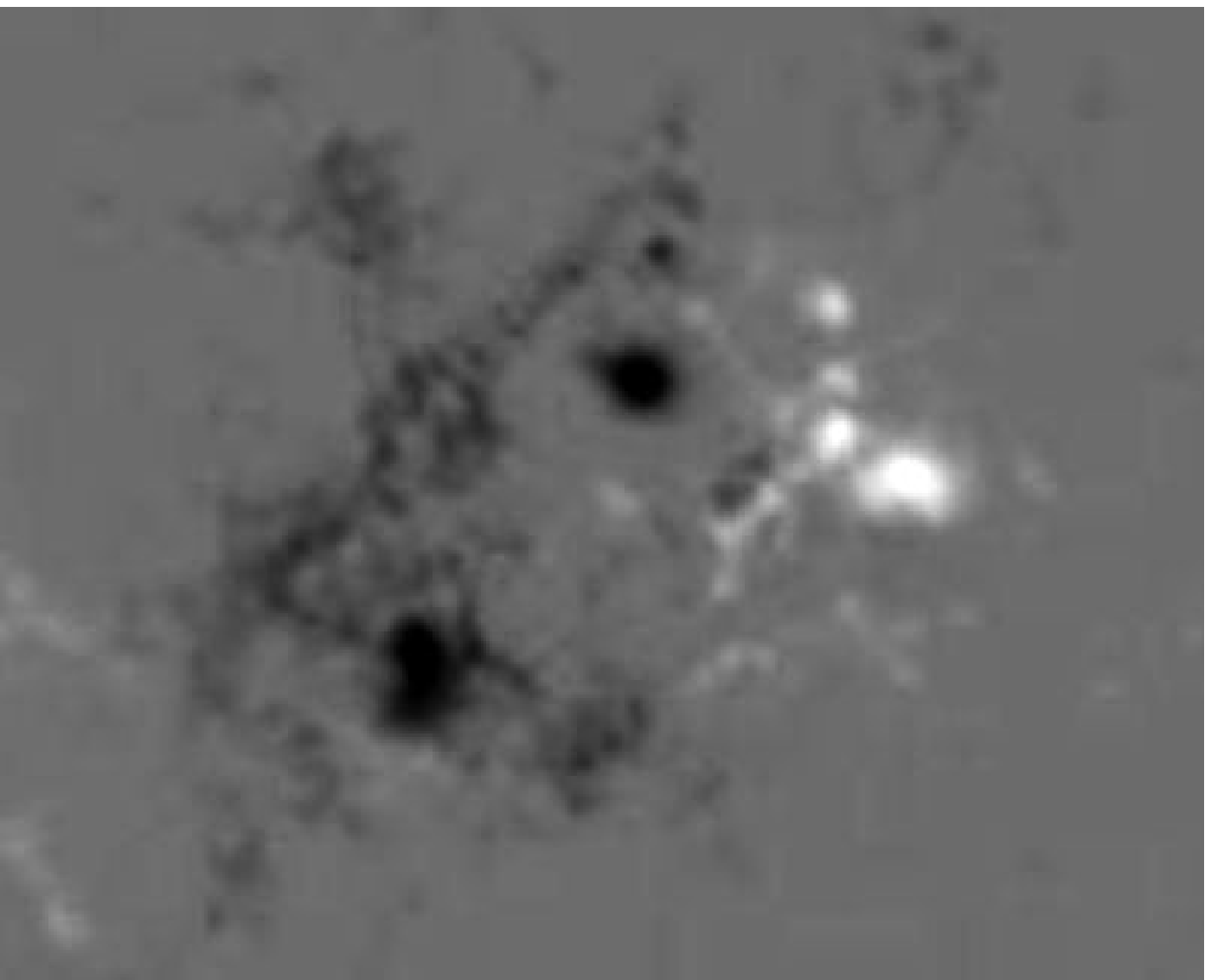}
\includegraphics[width=0.192\textwidth]{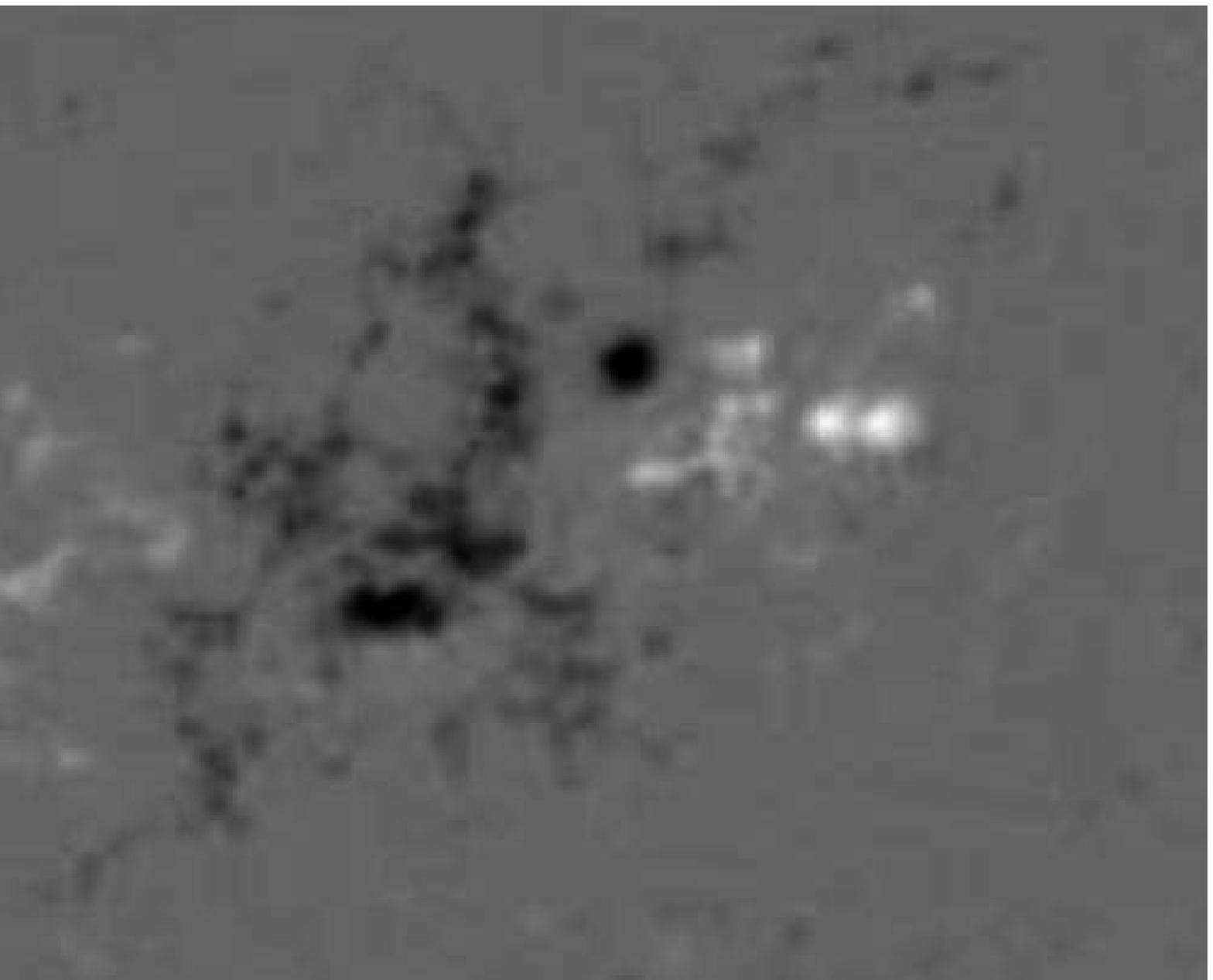}
}

\centerline{
\includegraphics[width=0.192\textwidth]{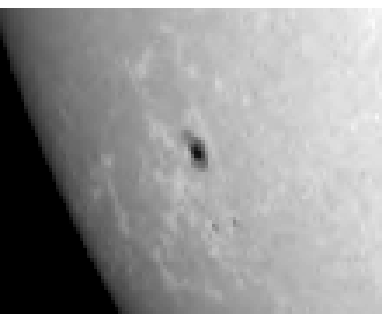}
\includegraphics[width=0.192\textwidth]{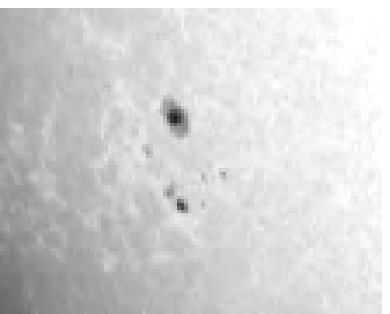}
\includegraphics[width=0.192\textwidth]{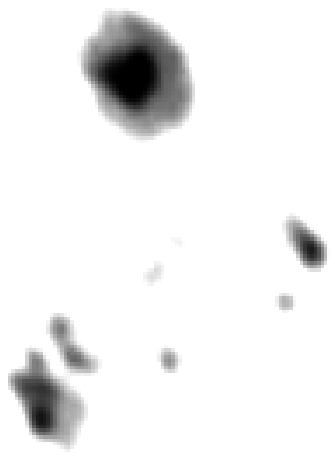}
\includegraphics[width=0.192\textwidth]{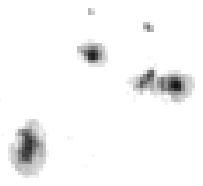}
\includegraphics[width=0.192\textwidth]{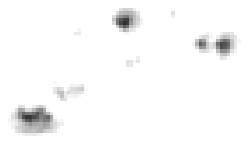}
}

\vspace{-0.215\textwidth}   
\centerline{\Large \bf     
\hspace{0.13\textwidth}   \color{white}{(a)}
\hspace{0.112\textwidth}  \color{white}{(c)}
\hspace{0.112\textwidth}  \color{white}{(e)}
\hspace{0.112\textwidth}   \color{white}{(g)}
\hspace{0.12\textwidth}  \color{white}{(i)}
\hfill}

\vspace{0.122\textwidth}   
\centerline{\Large \bf     
\hspace{0.13\textwidth}   \color{black}{(b)}
\hspace{0.112\textwidth}  \color{black}{(d)}
\hspace{0.112\textwidth}  \color{black}{(f)}
\hspace{0.112\textwidth}   \color{black}{(h)}
\hspace{0.12\textwidth}  \color{black}{(j)}
\hfill}

\caption{Evolution of AR NOAA 9264 (the South hemisphere) illustrated by MDI 
magnetograms (upper row) and continuum images (bottom row) on 11 December 
2000 (a,b), 12 (c,d), 13 (e,f), 15 (g,h), 17 (i,j). A new reverse-polarity 
bipole emerged near the old unipolar spot.}
\label{Fig_6}
\end{figure}

We did not include the above mentioned cases in the compiled catalog.

iv) Tilt instability of emerging ARs. The tilt is known to be 
unstable during several days after the emergence onset (see, e.g., 
\citealp{vanDriel15}; \citealp{Schunker20}) and may change the orientation
importantly. One cannot determine the leading/following sunspot position 
and polarity unambiguously while the tilt is not stabilized. 
So, it is a challenge to establish a reverse polarity for the groups 
arising in the close proximity to the western limb, as well as for small 
short-lived ARs. We discarded ARs emerging near the western limb and 
observed only for one or two days. All short-lived ARs (having spots/pores 
of both polarities for less than three days) and ARs with tilt uncertainties
were indicated with special marks: ``S'' and ``T'', respectively 
(see Section 3.2). 

v) Emergence of a new reverse-polarity bipole close to a pre-existing 
unipolar sunspot. Sometimes such new bipoles have an individual NOAA 
number, e.g. ARs NOAA 7062 close to 7056, and sometimes they do not.  
We extracted the new anti-Hale region from the whole activity complex 
(in the latter case, under the hosting NOAA number). In the catalog, 
they are marked by ``X''. A typical example is shown 
in Figure~\ref{Fig_6}. This procedure was applied only for emerging 
anti-Hale regions that survived longer than three days. Sunspot areas 
for extracted ARs were calculated by using a technique described 
in Section 3.2. 

\begin{figure}
\centerline{
\includegraphics[width=0.325\textwidth]{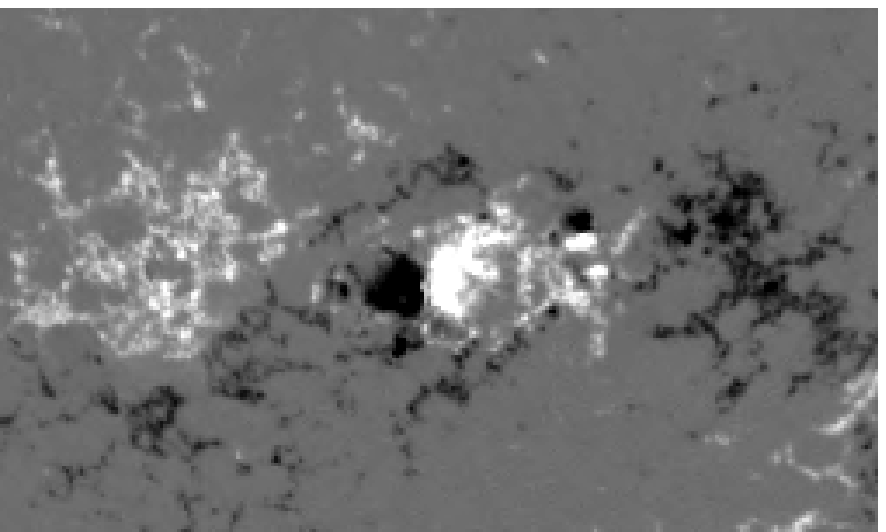}
\includegraphics[width=0.325\textwidth]{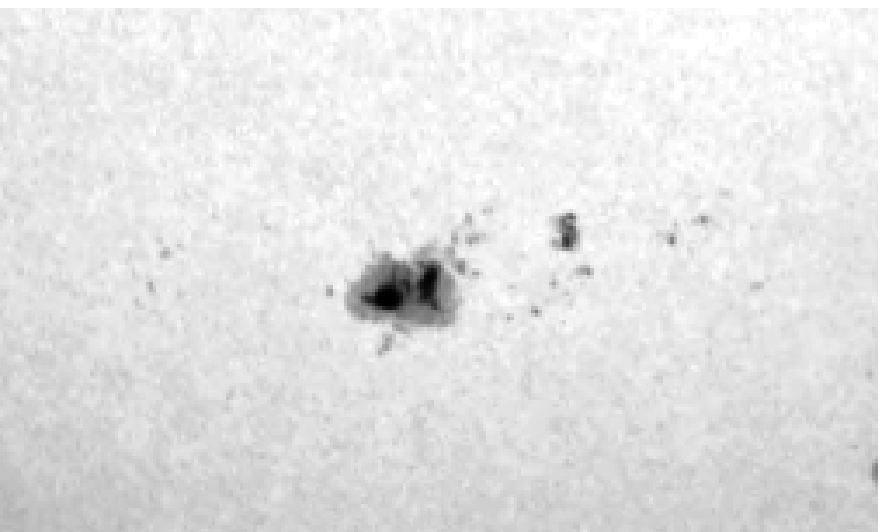}
\includegraphics[width=0.325\textwidth]{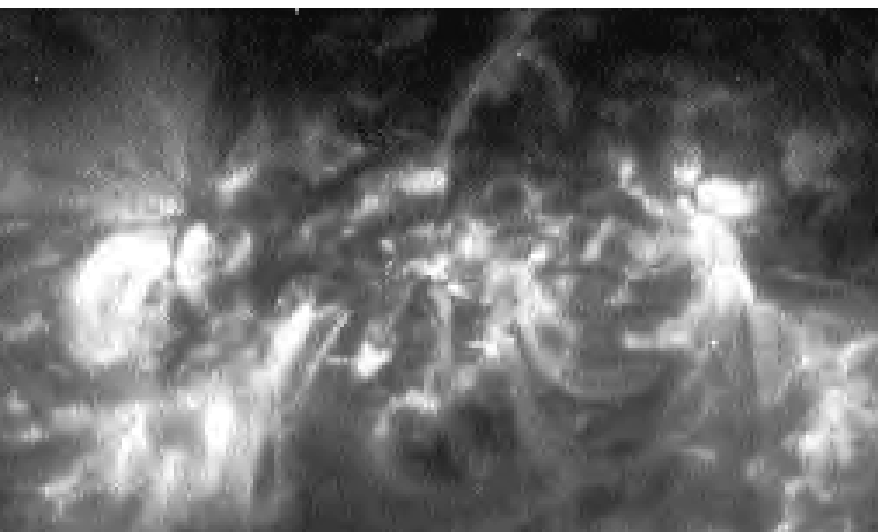}
}

\vspace{-0.052\textwidth}   
\centerline{\Large \bf     
\hspace{-0.008\textwidth}   \color{white}{(a)}
\hspace{0.245\textwidth}  \color{black}{(b)}
\hspace{0.248\textwidth}  \color{black}{(c)}
\hfill}

\caption{Complex AR NOAA 9591 including magnetic ``$\delta$-configuration'' 
sunspot observed on 28 August 2001 in the South hemisphere: MDI 
magnetogram (a), continuum (b), and AIA Fe {\sc ix/x} 171~\AA\ line (c) images. 
The ``$\delta$-sunspot'' of reverse polarity was extracted and considered 
in the compiled catalog.}
\label{Fig_7}
\end{figure}

vi) ARs changing their properties during the displacement across the disk. Some 
ARs, when passing across the disk, show anti-Hale properties for only a part 
of time. It may be due to emergence of a new magnetic flux in the vicinity. 
In other cases, a bipolar sunspot group with reverse-polarity 
remains after the decay of a multipolar AR. If the reverse-polarity property
is observed longer than three days, we noted this region with a special 
mark ``P''.  

vii) ARs with the main spots of both polarities near the Equator. 
\citet{McClintock14}
argue that equatorial location of sunspot groups is related 
to a misalignment of the magnetic and heliographic Equators. 
They suppose that ARs such as NOAA 11987, for example, pertain 
to the magnetic system of the North hemisphere despite the fact that 
the leading spot is located a bit below the Equator. The authors 
explain such phenomena by the different degree of activity of the North
and South hemispheres. Considering these reasons, we included into 
the catalog only those near-Equator ARs where the area-weighted center 
and the midpoint between the main spots of both polarities belong 
to the same hemisphere. They are marked with a letter ``E''.  

viii) ARs with ``$\delta$-configuration''. This class of ARs was introduced 
for sunspot groups with closely spaced opposite-polarity umbrae located 
in the common penumbra and added to the initial Mount Wilson magnetic 
classification by \citet{Kuenzel65}. A sample of AR NOAA 9591 including 
``$\delta$-sunspot'' is presented as Figure~\ref{Fig_7}. Individual
``$\delta$-structures'' with reverse polarity were included in our database 
with the mark ``D''. ``$\delta$-spots'' amidst the spread out background 
magnetic complex (see Figure~\ref{Fig_7}) were extracted and considered 
only in the cases when such a ``$\delta$-structure'' was the strongest 
feature in the entire complex. In case the extracted ``$\delta$-structure'' 
displays the reverse polarity, we indicated it with the mark ``W''. There are 
only seven such events in the compiled catalog.   

\begin{figure}
\centerline{
\includegraphics[width=0.192\textwidth]{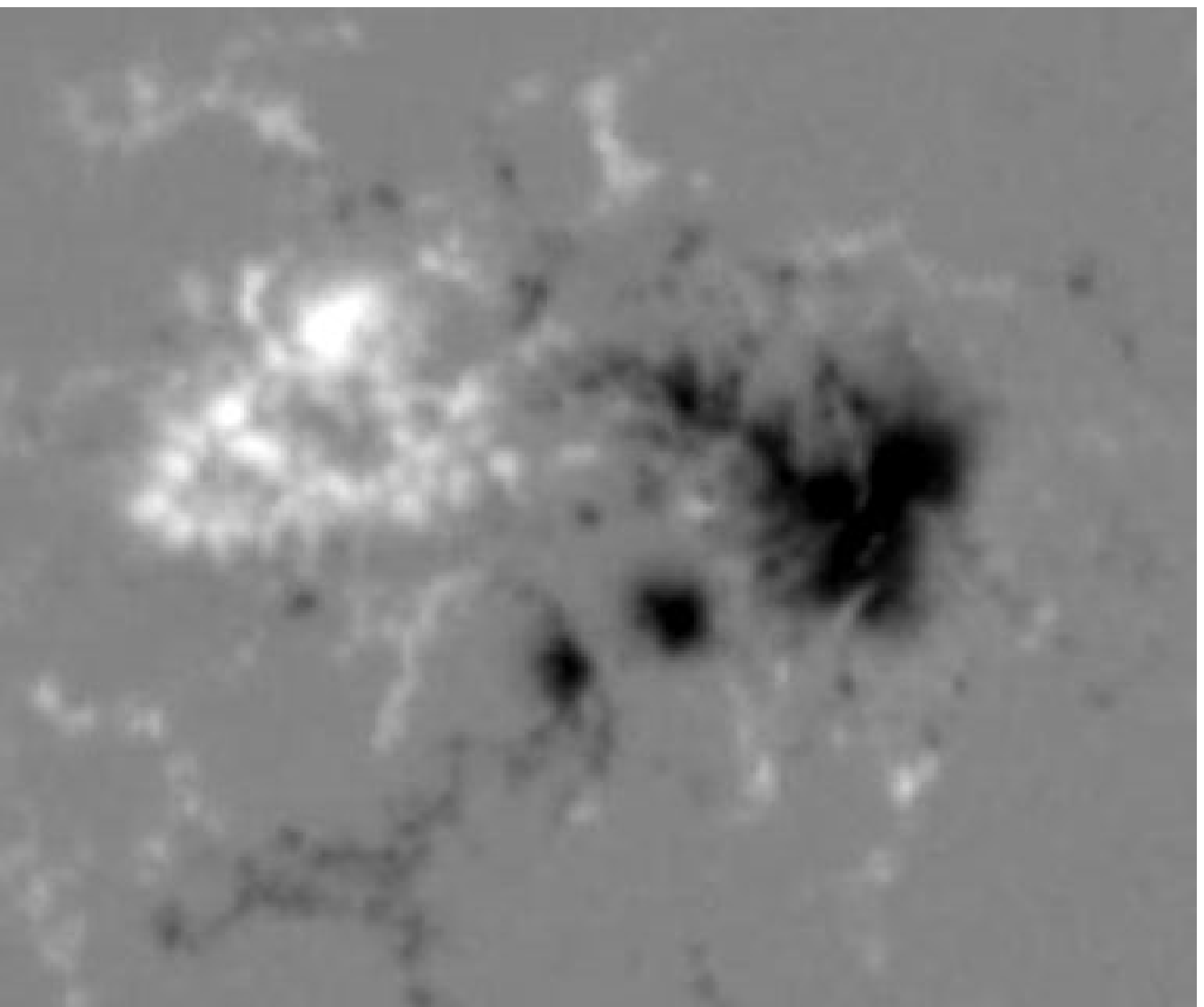}
\includegraphics[width=0.192\textwidth]{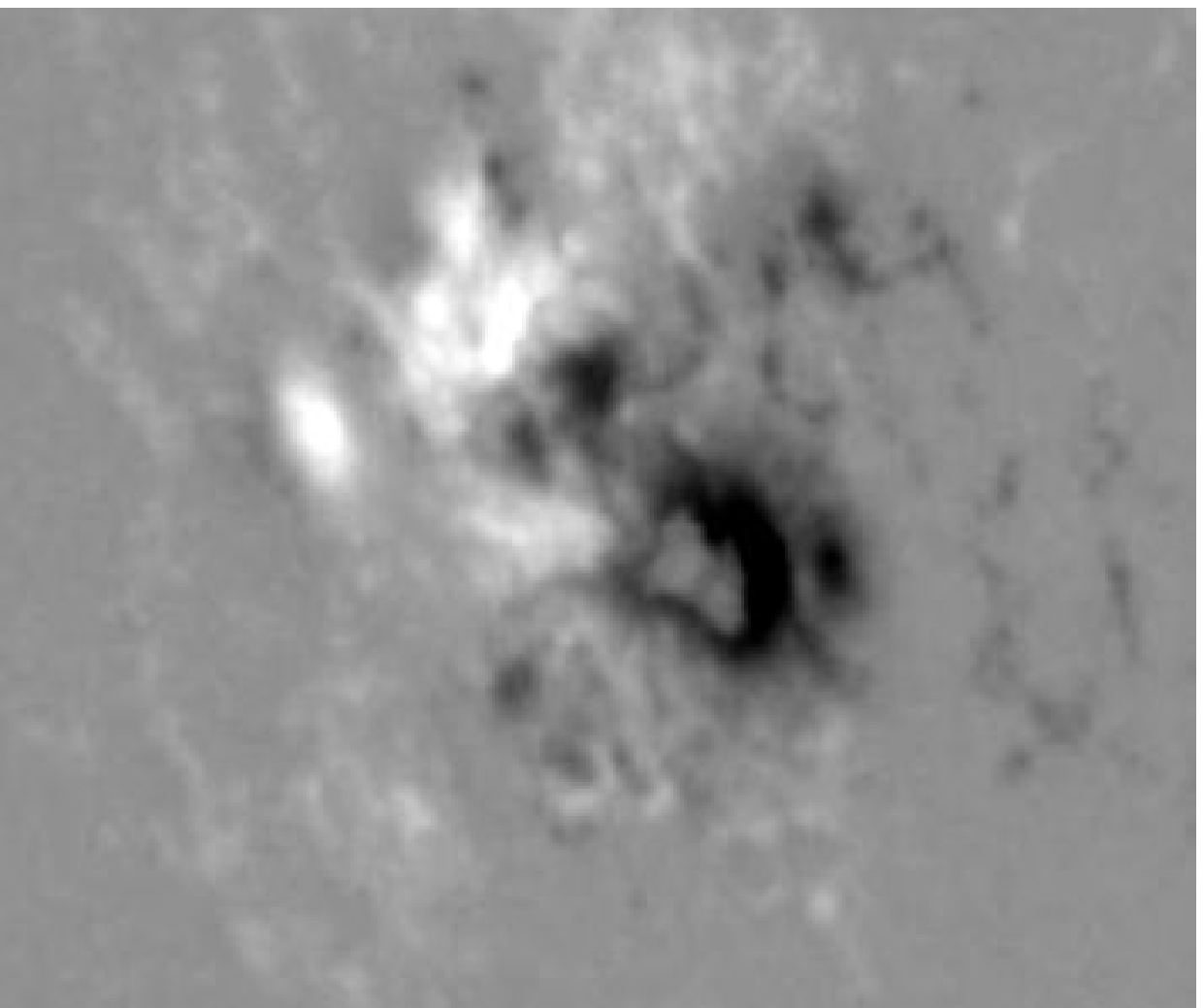}
\includegraphics[width=0.192\textwidth]{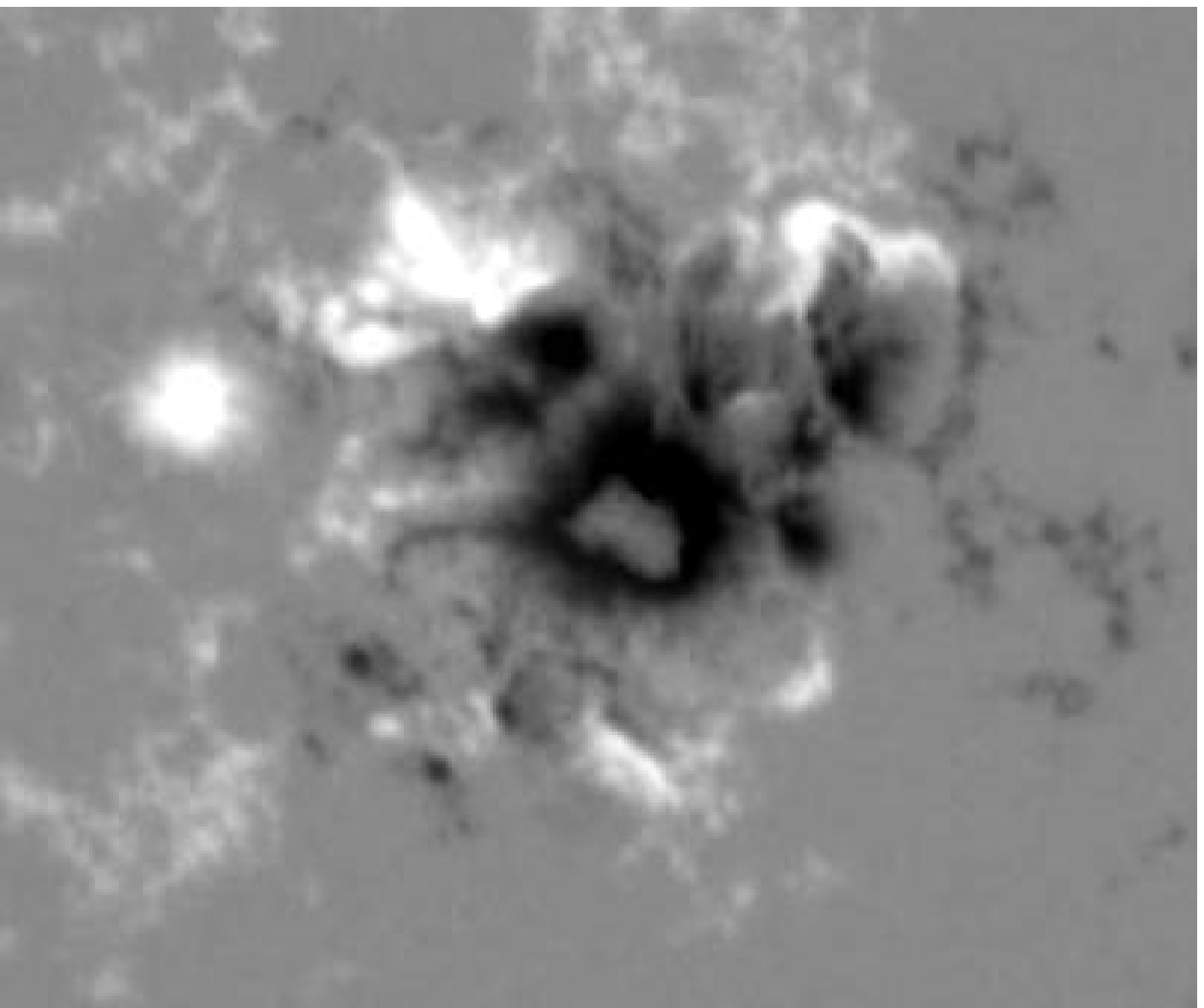}
\includegraphics[width=0.192\textwidth]{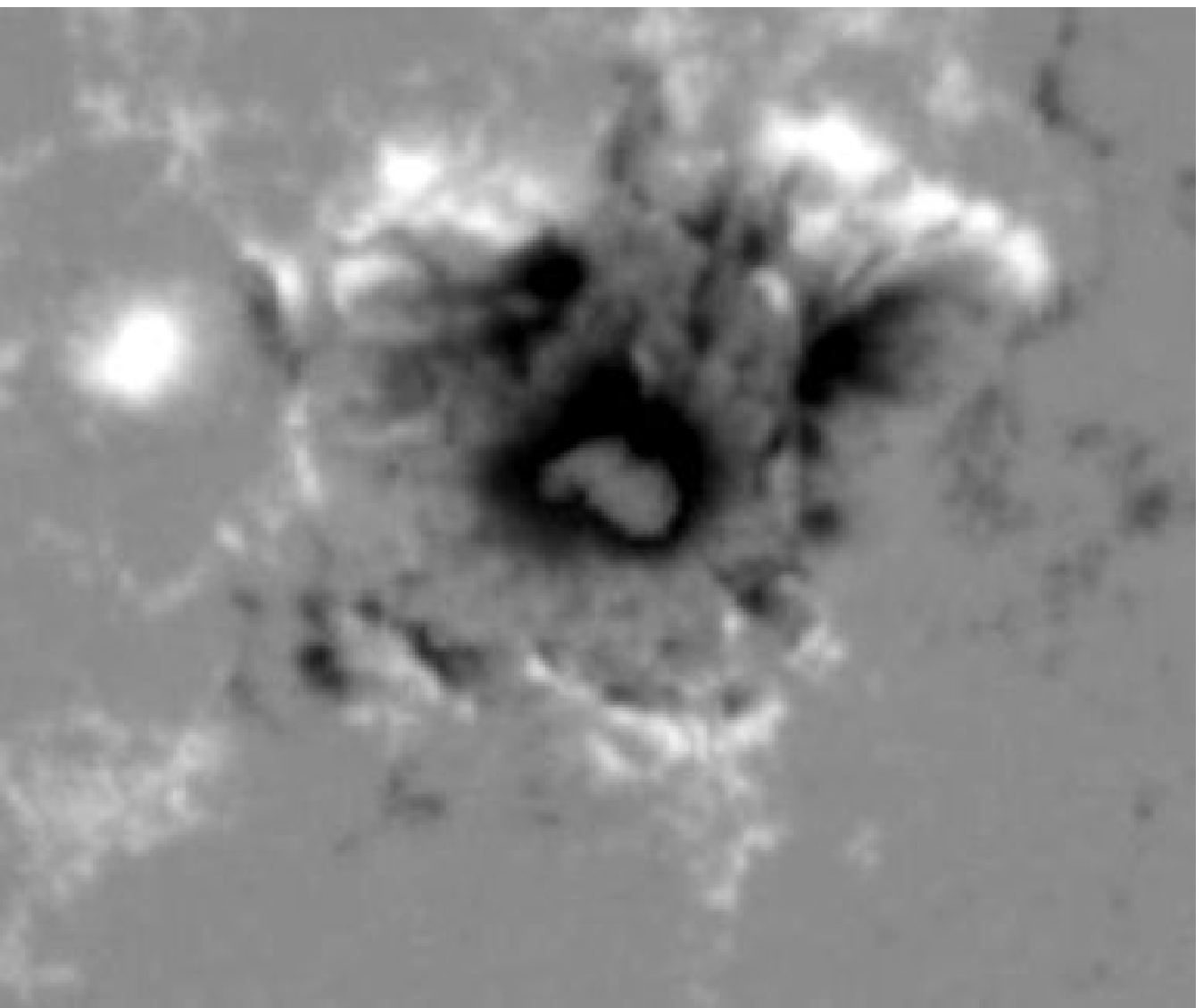}
\includegraphics[width=0.192\textwidth]{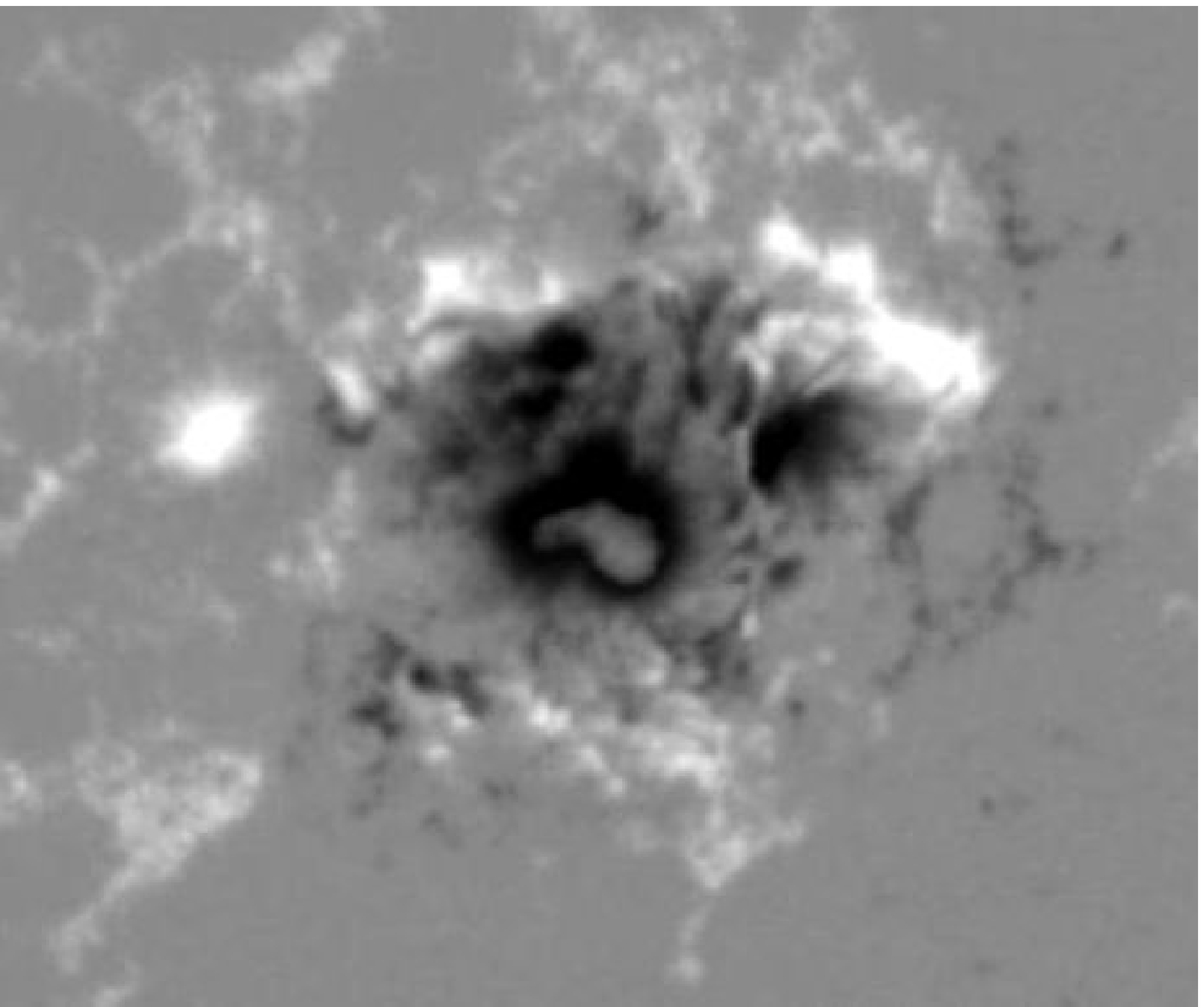}
}

\centerline{
\includegraphics[width=0.192\textwidth]{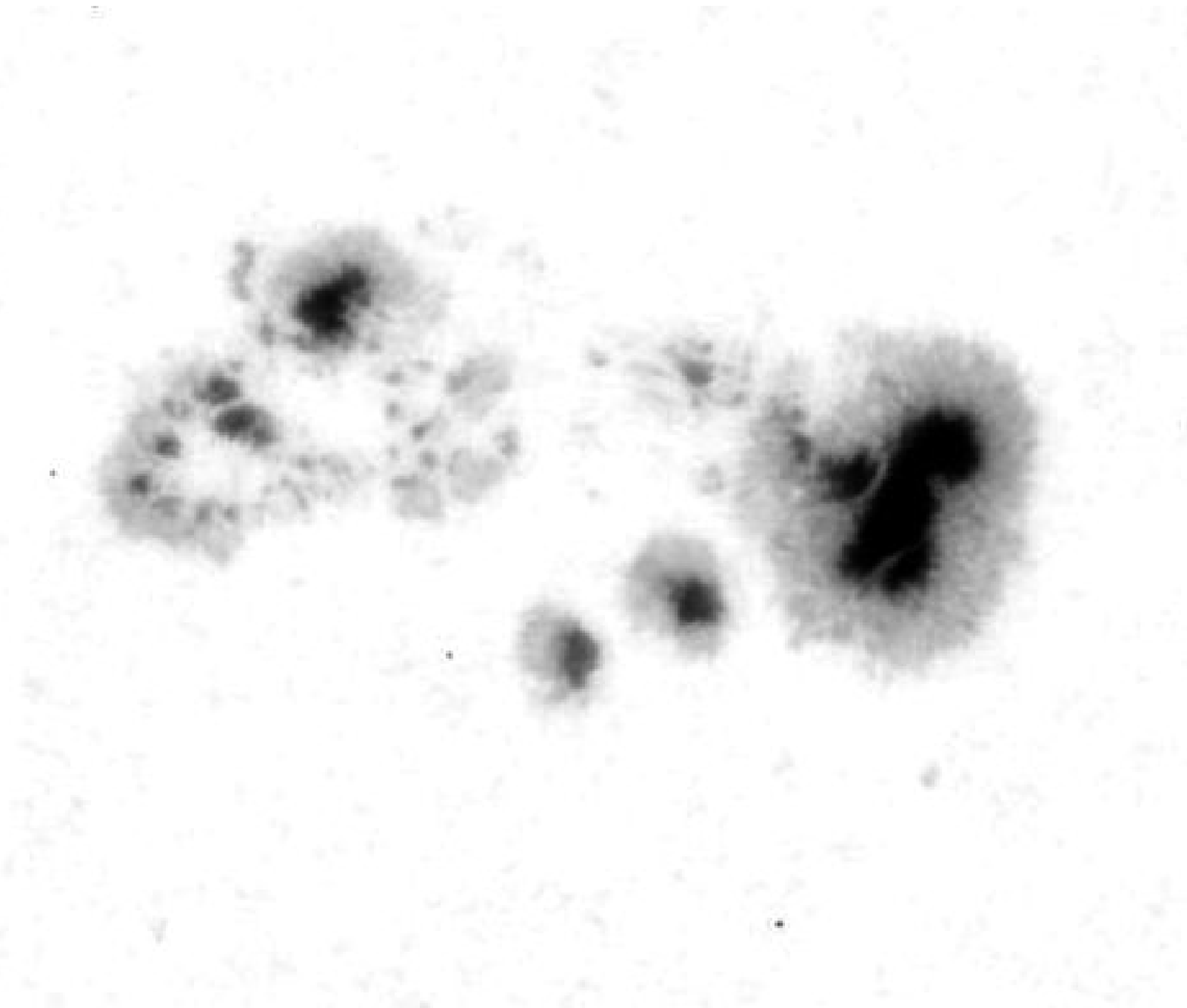}
\includegraphics[width=0.192\textwidth]{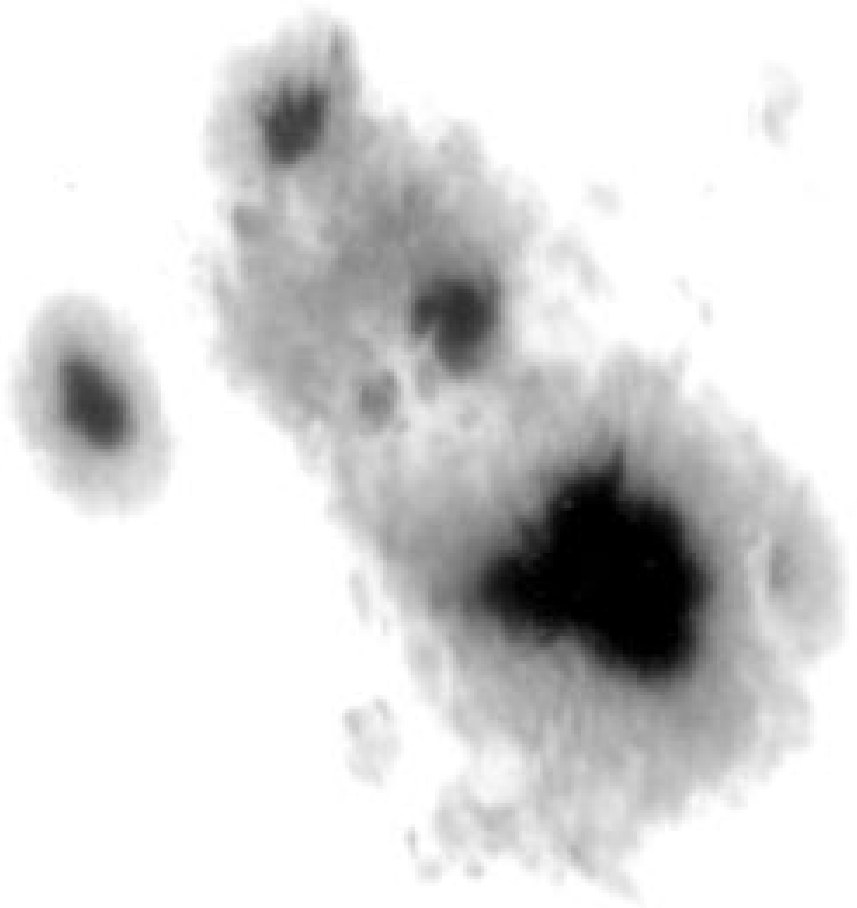}
\includegraphics[width=0.192\textwidth]{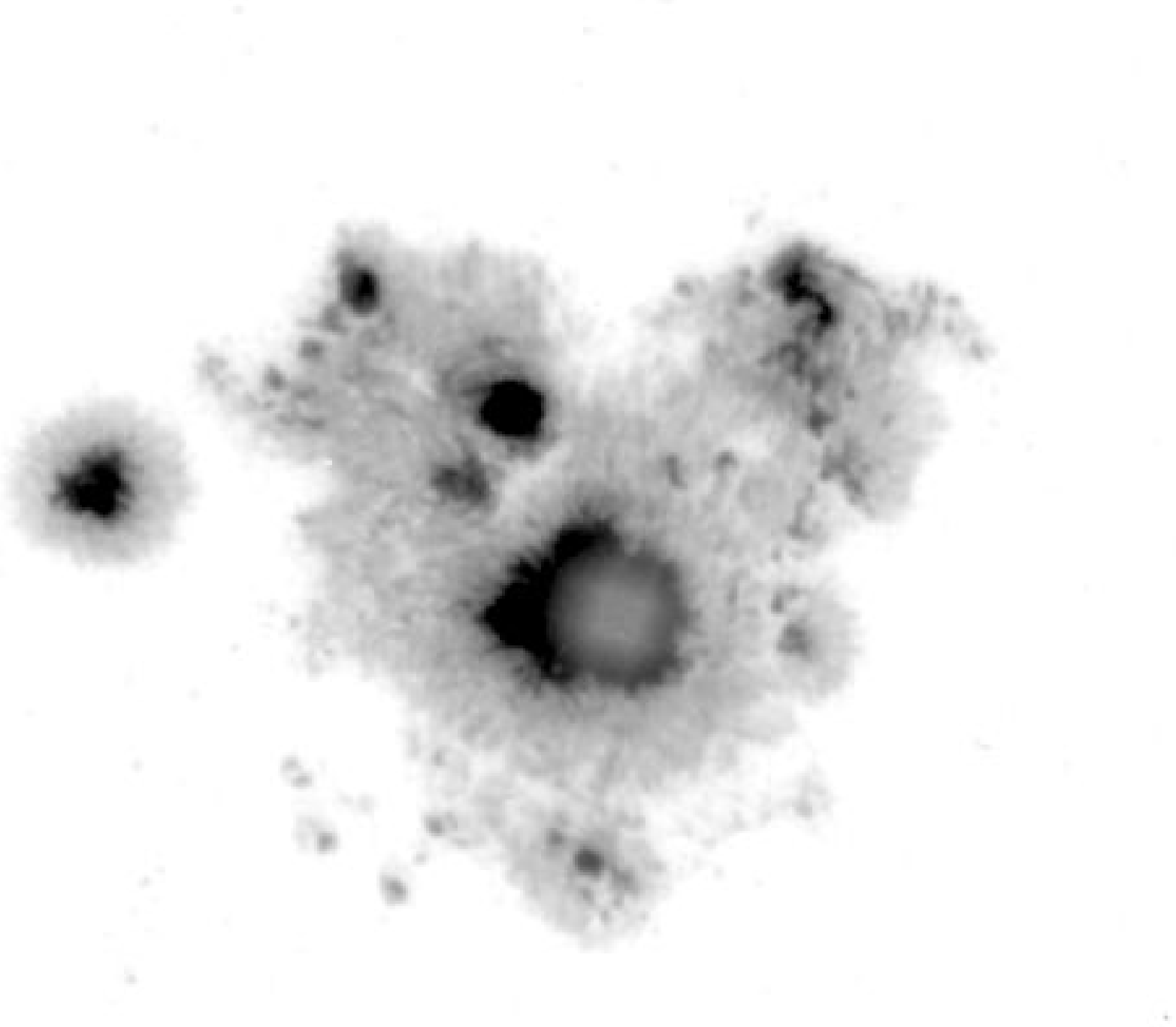}
\includegraphics[width=0.192\textwidth]{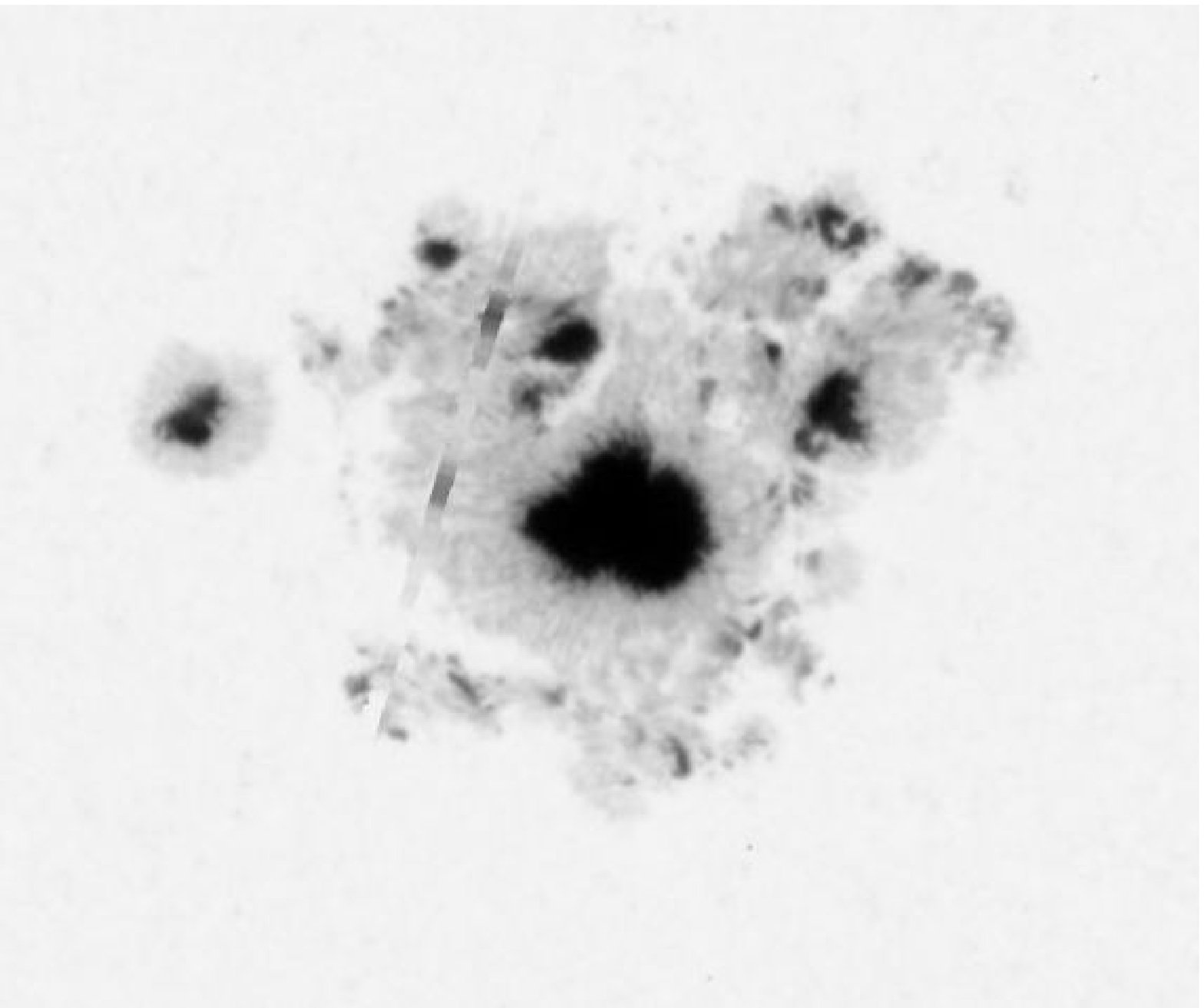}
\includegraphics[width=0.192\textwidth]{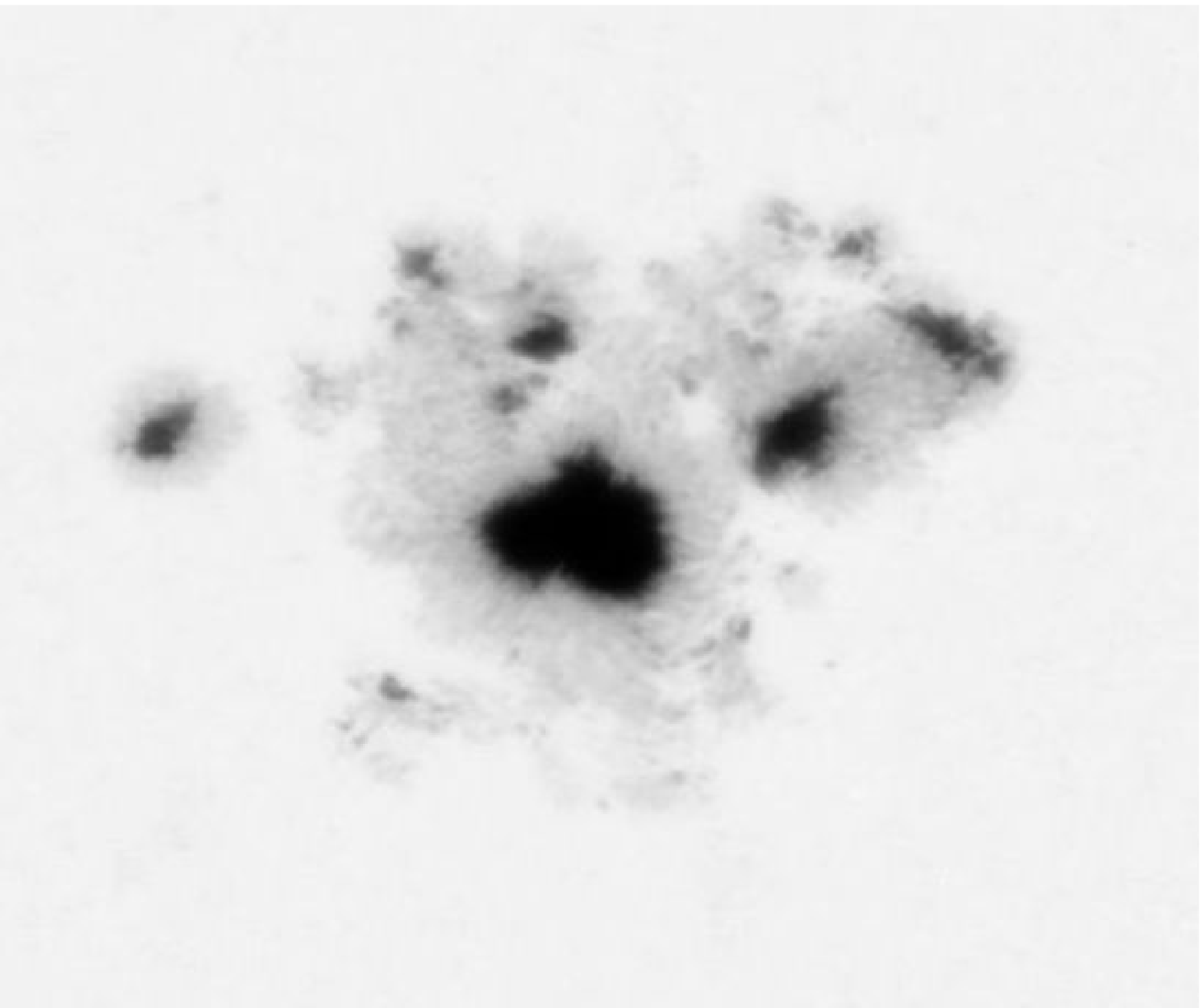}
}

\vspace{-0.215\textwidth}   
\centerline{\Large \bf     
\hspace{-0.01\textwidth}   \color{white}{(a)}
\hspace{0.112\textwidth}  \color{white}{(c)}
\hspace{0.112\textwidth}  \color{white}{(e)}
\hspace{0.112\textwidth}   \color{white}{(g)}
\hspace{0.12\textwidth}  \color{white}{(i)}
\hfill}

\vspace{0.122\textwidth}   
\centerline{\Large \bf     
\hspace{-0.01\textwidth}   \color{black}{(b)}
\hspace{0.112\textwidth}  \color{black}{(d)}
\hspace{0.112\textwidth}  \color{black}{(f)}
\hspace{0.112\textwidth}   \color{black}{(h)}
\hspace{0.12\textwidth}  \color{black}{(j)}
\hfill}

\caption{Recurrent AR NOAA 10036 (the South hemisphere) on 22 July 2002 (a,b) 
and its evolution as AR NOAA 10069 in the next rotation on 14 August 
2002 (c,d), 16 (e,f), 17 (g,h), 18 (i,j) illustrated by MDI magnetograms 
(upper row), DPD (Gyula Observing Station and Debrecen Heliophysical 
Observatory) continuum images (bottom row). The old leading sunspot is 
surrounded by opposite-polarity sunspots since 16 August; some of them are 
located at the western part of the AR. This one and similar multipolar ARs 
were not included in the catalog.}
\label{Fig_8}
\end{figure}

ix) Multipolar ARs. A sample of such AR (NOAA 10069) is shown 
in Figure~\ref{Fig_8}: a number of equivalent-by-size sunspots of both 
polarities are spread around without forming any single bipole. Sometimes 
the most western spots
may have the wrong polarity according to the Hale's polarity law, see 
Figure~\ref{Fig_8}. In spite of this attribute, we decided not to count
multipolar ARs in out dataset. Examples of multipolar ARs are: NOAA 6562, 
6624, 10501, 10786, and 12673.

x) The polarity of sunspots may be reversed due to the projection effect.
Of course, such cases were not counted.

\subsection{Shortcomings in the Existing Data Sources}

\begin{figure}
\centerline{
\includegraphics[width=0.315\textwidth]{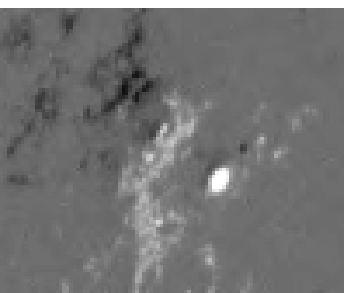}
\includegraphics[width=0.325\textwidth]{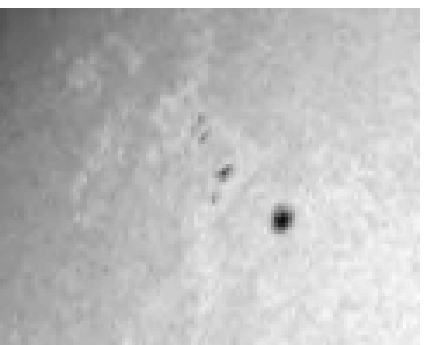}
\includegraphics[width=0.325\textwidth]{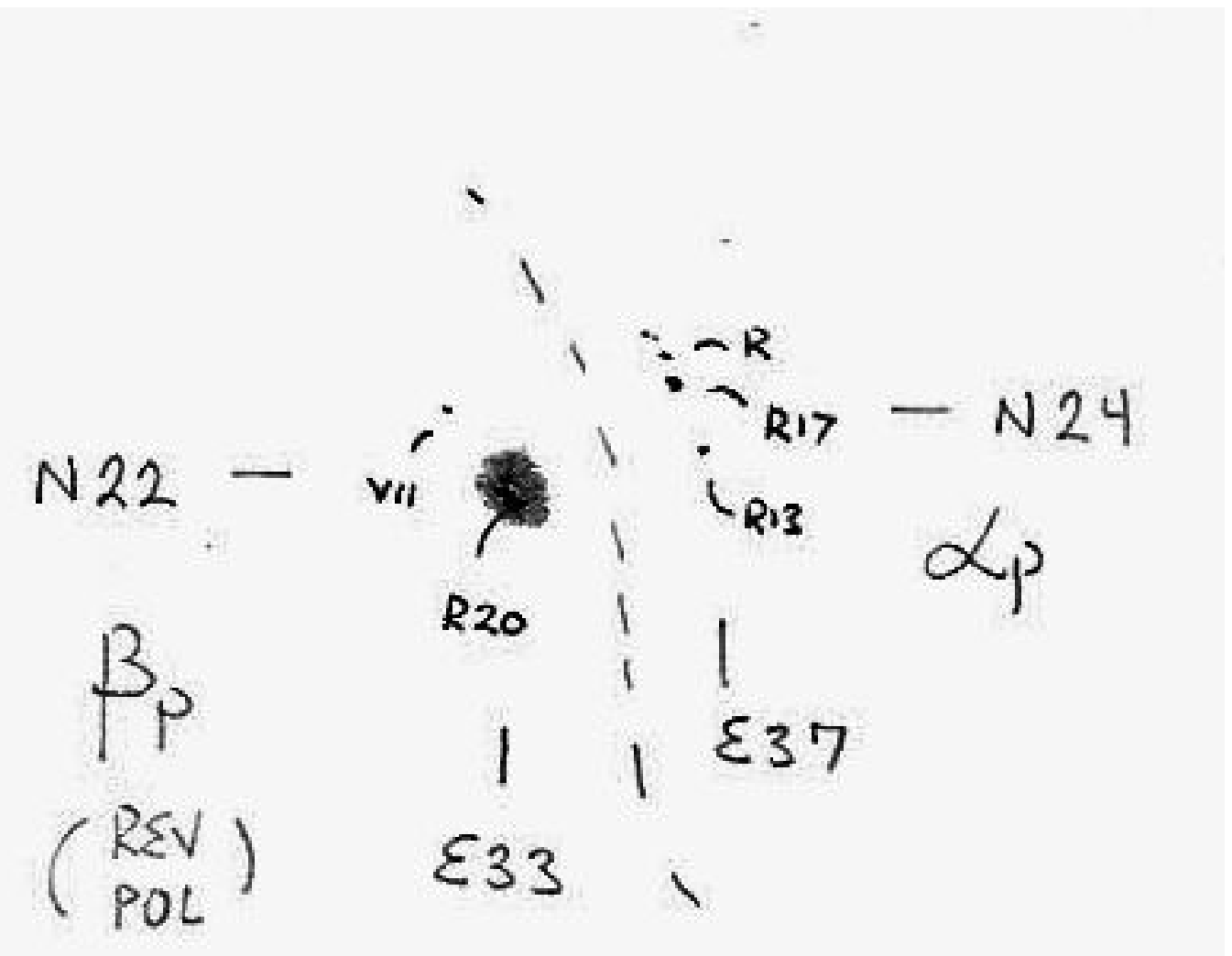}
}

\vspace{-0.052\textwidth}   
\centerline{\Large \bf     
\hspace{0.24\textwidth}   \color{white}{(a)}
\hspace{0.245\textwidth}  \color{black}{(b)}
\hspace{0.248\textwidth}  \color{black}{(c)}
\hfill}

\caption{AR NOAA 9294 observed on 04 January 2001 in the North hemisphere: MDI 
magnetogram (a) and continuum image (b). (c) an erroneous identification 
of the group as a group of reverse polarity on the MWO drawing.}
\label{Fig_9}
\end{figure}

We also encountered issues related to data sources. 

i) Flaws of the MWO data:

a) Lack of records (e.g. ARs NOAA 7926, 9858, and 10268) because of different
reasons, including weather conditions (e.g. ARs NOAA 6255, 7430,7872, and
8963).

b) Wrong determination of the magnetic-field polarity (all sunspots/pores 
of a bipolar AR are marked as the same polarity features, e.g. 
ARs NOAA 8023, 9470, and 9699). \citet{Wang15} also reports that such cases are 
multiple among ARs with the polar separation less than 2.5 degrees.

c) Erroneous usage or lack of the mark ``REV. POL.'' (reversed polarity) 
on MWO drawings. A sample of an AR with proper polarity of the leading 
sunspot, but wrongly marked as ``REV. POL.'' is presented 
in Figure~\ref{Fig_9}. 
The magnetogram and the continuum image have the usual EW-orientation 
whereas the MWO drawing has the opposite orientation. The main spot R20 
has a regular (for leading spots in Solar Cycle 23 in the North hemisphere) 
positive polarity, but the AR is marked as ``REV. POL.''. The opposite 
situation is observed in the MWO catalog for the anti-Hale ARs NOAA 5622, 
7741, 7948, 8210, 9045, 9736, and 10638, which are not marked 
as reverse-polarity ARs.

d) Lack of AR NOAA number.

ii) Uncertainties in AR NOAA assignement in the DPD, the USAF/NOAA SRS 
databases, Solar Monitor, and Helioviewer:

a) After the decay of an AR, the same NOAA number was assigned to a new 
anti-Hale region that emerged at the same location (e.g. ARs NOAA 6241, 7324) 
or nearby (e.g. AR NOAA 7062); 

b) Differences in outlining of boundaries (areas) of ARs 
in the USAF/NOAA SRS and the DPD databases. For example, according to 
the DPD, the anti-Hale AR NOAA 12310 is a single active region. But 
according to the USAF/NOAA SRS and Solar Monitor, this complex is fragmented 
into three different ARs (NOAA 12310, 12312, and 12313). Another example: 
AR NOAA 6203 is classified as a single group on the DPD sketch, 
and as two separate ARs (including the anti-Hale bipolar group MWO number 
26212) on the MWO drawing.

ARs with such ambiguities in various databases 
were indicated with mark ``N''. 

iii) Errors in calculations of sunspot-group areas. For instance, 
in the \\ USAF/NOAA SRS database on 13 July 1994 AR NOAA 7746 has 
an extremely 
high value of the group area, which triply exceeds that reported 
for previous and following days. At the same time, the DPD data show 
a smooth change of this value.  

To this end, to provide a database of anti-Hale regions without the
aforementioned shortcomings, we examined anew all ARs from 1989 to 2018.

\section{The Compiled Catalog}

\subsection{Anti-Hale Regions Selection Criteria}

When elaborating our criteria of identification of anti-Hale regions, 
we were guided by original ideas by \citet{Hale25}. We also assumed that 
the main spots of different polarities are footpoints of emerged magnetic 
loops in accordance with classic magnetic-cycle models (\citealp{Babcock61, 
 Parker55}). Following this, we formulated two basic 
criteria for identification of an AR as an anti-Hale region.

\begin{enumerate}
\item Dominating features of an AR form a bipole of reverse polarity 
with sunspots or pores of both polarities being present.

\item Magnetic connections between the opposite polarities are steady: 
magnetic loops (arcades) are visible at least in one of the EUV images 
day by day (in the cases when EUV data are available).

The remaining four additional criteria are proposed for certain ambiguous 
types of ARs.

\item For small, short-lived (less than three days) sunspot groups. 
The tilt of an AR has to be stable (not rotating, not perpendicular 
to the Equator). The rotation of the tilt
more than 20 degrees during the interval of observation was considered
as a non-stable tilt.
The area of an AR must be greater than 4 $\mu$Hem and the corrected 
umbra area by the DPD must be non-zero (to exclude ambiguous pores: 
``transient'' and ``penumbrae without 
umbrae'' features, see \citealt{Lefevre14}).

\item For sunspot groups located on the Equator. Both the area-weighted 
center and the midpoint between the polarities of an AR have to be located 
in the same (the North or South) hemisphere. 

\item For emerging bipolar groups appearing from a pre-existing AR. 
The extracted bipole has to exist at least three days.

\item For ``$\delta$-structures'' amidst the background magnetic complex. 
The extracted ``$\delta$-structure'' has to form the strongest feature 
in the magnetic complex and does not have steady magnetic connections
with other magnetic centers of the complex.
\end{enumerate}

We did not consider any unipolar sunspots. We also did not consider 
multipolar ARs. 

\subsection{Anti-Hale Database and Special Marks}

All detected anti-Hale ARs are listed in the database (see the Electronic 
Supplementary Material). Several lines from the catalog beginning, as well 
as the two last lines of the catalog are shown as Table \ref{Cat}. 
For each AR, the catalog includes three information blocks: 
i) the USAF/NOAA SRS data; 
ii) the DPD data; iii) our special marks. The first and second blocks each 
include the following parameters: the NOAA (or DPD) number; the date 
of the greatest corrected total area of sunspots in D M Y (day, month, year) 
format. For this date, we show: latitude (LAT), longitude (LON), 
the maximum value of the corrected total sunspots area (Area). Area values 
are provided in millionths of the solar hemisphere ($\mu$Hem).

\begin{table}
\caption{Catalog of the anti-Hale regions}
\label{Cat}
\Rotatebox{90}{
\begin{tabular}{@{}r@{}r@{}r@{ }l@{ }r@{ }r@{ }r@{\ }l@{}r@{}r@{ }l@{ }r@{ }r@{ }r@{\ }r@{ }c@{}c@{}c@{}c@{}c@{}c@{}c@{}l@{}}
\hline                 
\multicolumn{1}{c}{}  & \multicolumn{7}{c}{USAF/NOAA SRS} & \multicolumn{7}{c}{DPD} & \multicolumn{8}{c}{marks} \\
No. & \ NOAA & \,\,\  & D \, M \ Y  & LAT,  & LON,  & Area,    & Hale  & NOAA & \,\,\  & D \, M \ Y & LAT,   & LON,   & Area,    & Tilt,  &1&2&3&4&5&6&7&\ AC   \\
    &        & \,\,\  &             & [degr.] & [degr.] & [$\mu$Hem] & class & /DPD & \,\,\  &            & [degr.]  & [degr.]  & [$\mu$Hem] & [degr.]  & & & & & & & &\      \\
\hline                                                                              
  1 & 5409   & \,\,\  & 23 03 1989  &  18.0   &  14.0   &  990       & BGD   & 5409 & \,\,\  & 22 03 1989 &  16.79   &  12.46   & 1334       & -26.93   & & & & &P& &V&\ 1.00 \\
  2 & 5414   &        & 24 03 1989  & -18.0   &  49.0   &  150       & B     & 5414 &        & 21 03 1989 & -18.01   &  20.65   &  238       &  26.55   & & & & & & &V&\ 1.00 \\
  3 & 5441   &        & 10 04 1989  &  36.0   & -15.0   &  330       & BD    & 5441 &        & 08 04 1989 &  35.03   & -29.55   &  506       & -24.66   & & & & & & &V&\ 1.00 \\
  4 & 5455   &        & 14 04 1989  &  19.0   & -58.0   &   20       & B     & 5455 &        & 15 04 1989 &  24.93   & -31.74   &    6       & -17.10   &S& & & & & &V&\ 1.00 \\
  5 & 5470   &        & 05 05 1989  &  28.0   & -17.0   &  520       & BGD   & 5470 &        & 04 05 1989 &  29.44   & -16.97   &  775       &  15.82   & & & & & & &V&\ 1.00 \\  
  6 & 5504   &        & 24 05 1989  &  22.0   &  -9.0   &   20       & B     & 5504 &        & 23 05 1989 &  22.27   &  -7.07   &   17       &  21.93   &S& & & & & &V&\ 1.00 \\
  7 & 5548   &        & 17 06 1989  & -22.0   &  58.0   &   10       & A     & 5548 &        & 16 06 1989 & -21.55   &  57.96   &    5       & -47.77   &S& &T& & & &V&\ 1.00 \\
  8 & 5600   &        & 22 07 1989  & -18.0   & -35.0   &   10       & B     & 5600 &        & 22 07 1989 & -19.42   & -27.01   &   37       &  zzzzz   & & & & &X& &V&\ 0.52 \\
  9 & 5622   &        & 09 08 1989  & -26.0   &  44.0   &  280       & B     & 5622 &        & 08 08 1989 & -26.07   &  42.26   &  444       & -13.28   & & & & &P& &V&\ 1.00 \\
 10 & 5648   &        & 17 08 1989  &   7.0   &  14.0   &   10       & B     & 5648 &        & 17 08 1989 &   6.80   &  29.83   &    9       &  -5.91   &S& & & & &N&V&\ 1.00 \\
 11 & 5656   &        & 23 08 1989  &  26.0   & -32.0   &   10       & B     & 5656 &        & 23 08 1989 &  25.46   & -18.97   &   12       &   0.50   &S& & & & & &V&\ 1.00 \\
 12 & 5703   &        & 22 09 1989  & -23.0   & -45.0   &   60       & B     & 5703 &        & 21 09 1989 & -22.14   & -44.66   &   72       &  19.27   &C& & & &P& &V&\ 1.00 \\
 13 & 5729   &        & 06 10 1989  & -28.0   & -61.0   &   50       & B     & 5729 &        & 08 10 1989 & -28.25   & -24.43   &   49       &  -0.99   & & & & & & &V&\ 1.00 \\
 14 & 5747   &        & 19 10 1989  & -27.0   & -22.0   & 1160       & BD    & 5747 &        & 19 10 1989 & -26.72   & -12.04   & 1354       & -38.64   & &D&T&R&P& &V&\ 1.00 \\                      
 ...\\                                                                                                                                                                  
274 &12703   &        & 31 03 2018  & -8.0    & -60.0   &   10       & A     &12703 &        & 30 03 2018 & -9.17    & -71.68   &   24       & zzzzz    &C& & & & & & &\ 1.00 \\   
275 &12720   &        & 26 08 2018  &  8.0    &  39.0   &  100       & B     &zzzzz &        & zz zz zzzz &zzzzzz    & zzzzzz   & zzzz       & zzzzz    & & & & & & & &\ 1.00 \\    
\hline
\end{tabular}
}
\end{table}

The most complex Mount Wilson magnetic class (Hale class) achieved 
by a given AR is also shown in the USAF/NOAA SRS data block. When 
the Mount Wilson magnetic class was not available, we assigned this class. 
For ARs that exhibit the anti-Hale behaviour only for a partial time during 
the evolution, we considered only the anti-Hale period to determine the Mount 
Wilson magnetic class. 

The DPD block contains the tilt information for the ARs studied. 
Unfortunately, there is a lack of tilt data from 2014 to 2018 
in the DPD database (except for some months in 2017). The sign ``z'' stands 
for missing data. The data on the recurrent ARs are provided in the catalog 
for each rotation separately.

We use the following special marks for particular types of ARs:

C -- decaying ({\bf C}ollapsing) ARs;

D -- individual ``{\bf$ \delta$}-sunspots'';

E -- {\bf E}quatorial ARs;

N -- ARs with ambiguities in the
\textbf{N}OAA numbers assignment and/or outlining of boundaries  
of sunspot groups in various databases;

P -- an AR exhibits the anti-Hale property only {\bf P}art time 
when passing across the disk;

R -- ARs with {\bf R}otation of the sunspot-group axis;

S -- small {\bf S}hort-lived ARs;

T -- ARs with the {\bf T}ilt close to 90$^{\circ}$;

V -- ARs which are not checked for the stability of magnetic connections 
(lack of EU\textbf{V} data);

W -- ``$\delta$-sunspots'' {\bf W}ith surrounding sunspots/pores; 

X -- e{\bf X}tracted ARs.

The extracted ARs were counted with the hosting AR NOAA numbers. Area
values in such cases were determined as follows. For the DPD data block, 
the area of each extracted bipole at the evolutionary maximum was found 
as a total of composing sunspots. For the USAF/NOAA SRS data block 
(the individual sunspot-area data are absent), we used a different method. 
From the DPD data, we determined the moment of the maximal evolution 
for a given extracted anti-Hale region. Then, an auxiliary coefficient 
was found (also from the DPD data) as a ratio of the corrected total 
sunspot area of the extracted anti-Hale region to the corrected total 
sunspot area of the entire hosting group. This coefficient was applied 
to the area value of the hosting AR from the USAF/NOAA SRS database 
(at the same date) to calculate the anti-Hale region area.
The auxiliary coefficient marked as ``AC''
is shown in the last column of the catalog. 

Areas of ARs with shortcomings in areal data (see Section 2.3) were 
calculated by special procedure. For example, only the areas of relevant 
sunspots were counted for AR NOAA 12310 for the DPD data block.

\section{Estimations of the Relative Number of Anti-Hale Regions}

In total, we studied 8606 ARs of Solar Cycles 22, 23, and 24. 275 sunspot 
groups were identified as anti-Hale regions and included in the catalog. 
We found the anti-Hale percentage as $\approx$3.2, and $\approx$3.0 
in the case of excluding short-lived ARs with unsteady tilt. 
This estimation is close to previous earliest results by \citet{Hale25}, 
$\approx$2.4, and by \citet{Richardson48}, $\approx$3.1.   

However, as we show in the Introduction, the recent studies of anti-Hale 
regions provide a higher percentage of 4\,--\,8. Possible reasons for this are
discussed below.

We found that small short-lived groups (96 ARs) comprise about a one third  
of our set of anti-Hale regions. Exclusion of these ARs changes 
the anti-Hale percentage significantly (down to $\approx$2.0).
Since \citet{Richardson48}, it is well known that Hale's polarity-law 
violators happen more often for small ($<$50~$\mu$Hem) sunspot groups. 
The frequency of this violation increases rather steeply with transition 
to regions of smaller size (\citealp{Hagenaar01, Tlatov10, Stenflo12}). 
Therefore, including smaller and smaller groups into a sample causes 
increases the anti-Hale region percentage.

The sensitivity of modern instruments (\citealp{Scherrer95, Scherrer12}) 
allows researchers to involve in the statistical studies weaker and weaker 
ARs. To estimate the magnetic-flux threshold for ARs selection in our 
study, we evaluated fluxes of several small short-lived ARs from 
the catalog. We found that the total unsigned magnetic-flux value 
of short-lived ARs having at least pores of both polarities (according 
to our criteria) is $\approx$$10^{21}$\,Mx. This allows us to consider this 
flux value as an indirect estimate of the threshold for our selection. 
We presume that counting of ARs with smaller magnetic fluxes is one 
of the main reasons for the enhanced percentage of anti-Hale groups 
in recent studies.

Another reason for the discrepancy might be the numerous ambiguities 
in anti-Hale region identification mentioned in Section 2.2. 
For example, the detection of anti-Hale ARs in activity complexes is very 
challenging (Section 2.2, item 2). According to data 
by \citet{Yazev11, Yazev15}, there are 321 activity complexes in Solar 
Cycles 22\,--\,24, which corresponds 900\,--\,1100 ARs (9\,--\,12\,\% 
from all ARs). Using average evaluation of anti-Hale ARs percentage 
$\approx$4.5 (see the Introduction), we roughly estimate a possible number 
of anti-Hale regions in activity complexes for three mentioned cycles 
as 40\,--\,50 ARs (up to 0.5\,\% from all ARs). This is an upper limit 
of anti-Hale ARs number which might be missed in our catalog.

We also compared the anti-Hale regions list for 1989\,--\,2004 that 
was used by \citet{Sokoloff15} (the 2015 database) with the co-temporal 
fragment of the compiled catalog. We found that some changes occured 
in the list of anti-Hale regions; the number of ARs in both databases 
remains close. The changes turned out to be evenly distributed along 
the cycle, so that the main inference made by \citet{Sokoloff15} about 
the cycle variations of the relative number of anti-Hale groups 
did not change.

\section{Conclusions} 

To identify ARs violating the Hale's polarity law, we visually examined 
sunspot groups from 1 January 1989 to 31 December 2018. This period 
encompasses the end of the Solar Cycle 22 and the two subsequent cycles 
including two solar minima, which is important for understanding 
the coupling between local and global magnetic fields in terms of solar 
dynamo theory. We used the SOHO/MDI and the SDO/HMI data, as well as 
the DPD sketches and the MWO catalog and drawings. Detailed comparison 
of data from different source allowed us to detect a number of ambiguous 
cases in anti-Hale sunspot groups identification. We analysed the revealed 
errors and inaccuracies to elaborate the criteria for identification 
of sunspot groups as anti-Hale ARs. We can conclude the following.

\begin{enumerate}
\item The main reasons for possible errors in anti-Hale region 
identification are: 
\begin{enumerate}
\item incorrect determination of AR boundaries;
\item wrongly distributing of ARs between consecutive cycles during
solar minima;
\item presence of small sunspots/pores with an opposite (to a proper 
leading polarity by the Hale's law) polarity to the West from the 
old leading spots in decaying ARs; 
\item emergence of a new magnetic flux within decaying ARs; 
\item the wrong polarity of the most western spots in multipolar ARs;
\item location of an AR on the Equator;
\item appearance of the wrong polarity of sunspots located near the limb 
the projection effect).
\end{enumerate}

\item Two basic and four additional criteria for identifying bipolar
sunspot groups as anti-Hale regions are formulated. Our basic criteria
are based on the classical ideas of \citet{Hale25} and \citet{Babcock61}.
Namely, we assume that for an AR to be classified as an anti-Hale region,
its dominating features have to form a bipole of reverse polarity 
with sunspots/pores of both polarities being present. Magnetic connection 
between the opposite polarities has to be observed.

\item The catalog of anti-Hale regions for the period of our study 
is compiled. For each anti-Hale region, the catalog contains: 
NOAA (or DPD) number; date, coordinates, and corrected total sunspot 
area at the evolutionary maximum as determined from the DPD and 
the USAF/NOAA SRS data. The catalog also contains the tilt and 
Mount Wilson magnetic class. In total, the catalog contains 275 ARs, 
including 96 small short-lived sunspot groups.

\item The percentage of sunspot groups violating the Hale's polarity law 
(meeting the formulated criteria) is $\approx$3.0 from all studied ARs. 
This is close to the earliest estimations by the authors who had examined 
each AR individually: $\approx$2.4 by \citet{Hale25} and $\approx$3.1 
by \citet{Richardson48}. Later researchers reported a slightly 
higher percentage. The enchancement of the percentage might be related
to increasing sensitivity of instruments and, therefore, taking into
account smaller and smaller bipoles. Another reason for the discrepancy 
in anti-Hale percentage is the ambiguities in the anti-Hale regions 
identification mentioned in Section 2.2.
\end{enumerate}

The catalog of anti-Hale regions can be used in studying the phenomena 
related to the cyclic solar activity. The catalog is available as 
the Electronic Supplementary Material. The catalog is also available 
at the CrAO website sun.crao.ru/databases/catalog-anti-hale.

\begin{acks}
 The authors are thankful to Drs. L.V. Ermakova and A.S. Kutsenko 
for consulting on complex, ambigous ARs. Efforts of A.V.Zhukova 
on the catalog compilation were supported by the RSF (Project 18-12-00131). 
A.I.Khlystova thanks the support of the RFBR's grants 18-02-00085 and 
19-52-45002. The work of A.I.Khlystova was supported by the basic 
financial program of the FSR II.16. D.D.Sokoloff is grateful for the RFBR 
support under the grant 18-02-00085. V.I.Abramenko would like 
to acknowledge partial support of the MSHE of RF (Research 0831-2019-0006).  
\end{acks}

\section*{Disclosure of Potential Conflicts of Interest}

The authors declare that they have no conflicts of interest.



\bibliographystyle{spr-mp-sola}
\bibliography{Zhukova_bibl}

\begin{thebibliography}{49}
\ifx\bisbn     \undefined \def\bisbn  #1{ISBN #1}\fi
\ifx\binits    \undefined \def\binits#1{#1}\fi
\ifx\bauthor   \undefined \def\bauthor#1{#1}\fi
\ifx\batitle   \undefined \def\batitle#1{#1}\fi
\ifx\bjtitle   \undefined \def\bjtitle#1{\textit{#1}}\fi
\ifx\bvolume   \undefined \def\bvolume#1{\textbf{#1}}\fi
\ifx\byear     \undefined \def\byear#1{#1}\fi
\ifx\bissue    \undefined \def\bissue#1{#1}\fi
\ifx\bfpage    \undefined \def\bfpage#1{#1}\fi
\ifx\blpage    \undefined \def\blpage #1{#1}\fi
\ifx\burl      \undefined \def\burl#1{\textsf{#1}}\fi
\ifx\href      \undefined \def\href#1#2{\textsf{#2}}\fi
\ifx\betal     \undefined \def\betal{\textit{et al.}}\fi
\ifx\bctitle   \undefined \def\bctitle#1{#1}\fi
\ifx\beditor   \undefined \def\beditor#1{#1}\fi
\ifx\bbtitle   \undefined \def\bbtitle#1{\textit{#1}}\fi
\ifx\bedition  \undefined \def\bedition#1{#1}\fi
\ifx\bseriesno \undefined \def\bseriesno#1{\textbf{#1}}\fi
\ifx\blocation \undefined \def\blocation#1{#1}\fi
\ifx\bsertitle \undefined \def\bsertitle#1{\textit{#1}}\fi
\ifx\bsnm      \undefined \def\bsnm#1{#1}\fi
\ifx\bsuffix   \undefined \def\bsuffix#1{#1}\fi
\ifx\bparticle \undefined \def\bparticle#1{#1}\fi
\ifx\barticle  \undefined \def\barticle#1{}\fi
\ifx\binstitute  \undefined \def\binstitute#1{#1}\fi
\ifx\bpublisher  \undefined \def\bpublisher#1{#1}\fi
\ifx\doiurl    \undefined \def\doiurl#1{\href{#1}{\textsf{DOI}}}\fi
\makeatletter
\def\safeHref#1#2#3{\in@{http}{#2}\ifin@\href{#2}{#3}\else\href{#1#2}{#3}\fi}
\makeatother
\ifx\adsurl    \undefined
  \def\adsurl#1{\safeHref{https://ui.adsabs.harvard.edu/abs/}{#1}{\textsf{ADS}}}\fi
\ifx\arxivurl  \undefined
  \def\arxivurl#1{\safeHref{http://arxiv.org/abs/}{#1}{\textsf{arXiv}}}\fi
\ifx\botherref \undefined \def\botherref#1{}\fi
\ifx\url       \undefined \def\url#1{\textsf{#1}}\fi
\ifx\bchapter  \undefined \def\bchapter#1{}\fi
\ifx\bbook     \undefined \def\bbook#1{}\fi
\ifx\bcomment  \undefined \def\bcomment#1{#1}\fi
\ifx\oauthor   \undefined \def\oauthor#1{#1}\fi
\ifx\citeauthoryear \undefined\def \citeauthoryear#1{#1}\fi
\def\endbibitem {}
\ifx\bconflocation  \undefined \def\bconflocation#1{#1} \fi

\bibitem[\protect\citeauthoryear{{Abramenko}, {Zhukova}, and
  {Kutsenko}}{2018}]{Abramenko18}
\begin{barticle}
\bauthor{\bsnm{{Abramenko}}, \binits{V.I.}},
\bauthor{\bsnm{{Zhukova}}, \binits{A.V.}},
\bauthor{\bsnm{{Kutsenko}}, \binits{A.S.}}:
\byear{2018},
\batitle{{Contributions from Different-Type Active Regions Into the Total Solar
  Unsigned Magnetic Flux}}.
\bjtitle{Geomag. Aeron.}
\bvolume{58},
\bfpage{1159}.
\doiurl{https://doi.org/10.1134/S0016793218080224}.
\adsurl{2018Ge&Ae..58.1159A}.
\end{barticle}
\endbibitem

\bibitem[\protect\citeauthoryear{{Babcock}}{1961}]{Babcock61}
\begin{barticle}
\bauthor{\bsnm{{Babcock}}, \binits{H.W.}}:
\byear{1961},
\batitle{{The Topology of the Sun's Magnetic Field and the 22-YEAR Cycle.}}
\bjtitle{\apj}
\bvolume{133},
\bfpage{572}.
\doiurl{https://doi.org/10.1086/147060}.
\adsurl{1961ApJ...133..572B}.
\end{barticle}
\endbibitem

\bibitem[\protect\citeauthoryear{{Baranyi}, {Gy{\H{o}}ri}, and
  {Ludm{\'a}ny}}{2016}]{Baranyi16}
\begin{barticle}
\bauthor{\bsnm{{Baranyi}}, \binits{T.}},
\bauthor{\bsnm{{Gy{\H{o}}ri}}, \binits{L.}},
\bauthor{\bsnm{{Ludm{\'a}ny}}, \binits{A.}}:
\byear{2016},
\batitle{{On-line Tools for Solar Data Compiled at the Debrecen Observatory and
  Their Extensions with the Greenwich Sunspot Data}}.
\bjtitle{\solphys}
\bvolume{291},
\bfpage{3081}.
\doiurl{https://doi.org/10.1007/s11207-016-0930-1}.
\adsurl{2016SoPh..291.3081B}.
\end{barticle}
\endbibitem

\bibitem[\protect\citeauthoryear{{Blackman} and {Field}}{2002}]{Blackman02}
\begin{barticle}
\bauthor{\bsnm{{Blackman}}, \binits{E.G.}},
\bauthor{\bsnm{{Field}}, \binits{G.B.}}:
\byear{2002},
\batitle{{New Dynamical Mean-Field Dynamo Theory and Closure Approach}}.
\bjtitle{Phys. Rev. Lett.}
\bvolume{89},
\bfpage{265007}.
\doiurl{https://doi.org/10.1103/PhysRevLett.89.265007}.
\adsurl{2002PhRvL..89z5007B}.
\end{barticle}
\endbibitem

\bibitem[\protect\citeauthoryear{{Brandenburg} and
  {Subramanian}}{2005}]{Brandenburg05}
\begin{barticle}
\bauthor{\bsnm{{Brandenburg}}, \binits{A.}},
\bauthor{\bsnm{{Subramanian}}, \binits{K.}}:
\byear{2005},
\batitle{{Astrophysical magnetic fields and nonlinear dynamo theory}}.
\bjtitle{Phys. Rep.}
\bvolume{417},
\bfpage{1}.
\doiurl{https://doi.org/10.1016/j.physrep.2005.06.005}.
\adsurl{2005PhR...417....1B}.
\end{barticle}
\endbibitem

\bibitem[\protect\citeauthoryear{{Brandenburg}, {Sokoloff}, and
  {Subramanian}}{2012}]{Brandenburg12}
\begin{barticle}
\bauthor{\bsnm{{Brandenburg}}, \binits{A.}},
\bauthor{\bsnm{{Sokoloff}}, \binits{D.}},
\bauthor{\bsnm{{Subramanian}}, \binits{K.}}:
\byear{2012},
\batitle{{Current Status of Turbulent Dynamo Theory. From Large-Scale to
  Small-Scale Dynamos}}.
\bjtitle{\ssr}
\bvolume{169},
\bfpage{123}.
\doiurl{https://doi.org/10.1007/s11214-012-9909-x}.
\adsurl{2012SSRv..169..123B}.
\end{barticle}
\endbibitem

\bibitem[\protect\citeauthoryear{{Hagenaar}}{2001}]{Hagenaar01}
\begin{barticle}
\bauthor{\bsnm{{Hagenaar}}, \binits{H.J.}}:
\byear{2001},
\batitle{{Ephemeral Regions on a Sequence of Full-Disk Michelson Doppler Imager
  Magnetograms}}.
\bjtitle{\apj}
\bvolume{555},
\bfpage{448}.
\doiurl{https://doi.org/10.1086/321448}.
\adsurl{2001ApJ...555..448H}.
\end{barticle}
\endbibitem

\bibitem[\protect\citeauthoryear{{Hale} and {Nicholson}}{1925}]{Hale25}
\begin{barticle}
\bauthor{\bsnm{{Hale}}, \binits{G.E.}},
\bauthor{\bsnm{{Nicholson}}, \binits{S.B.}}:
\byear{1925},
\batitle{{The Law of Sun-Spot Polarity}}.
\bjtitle{\apj}
\bvolume{62},
\bfpage{270}.
\doiurl{https://doi.org/10.1086/142933}.
\adsurl{1925ApJ....62..270H}.
\end{barticle}
\endbibitem

\bibitem[\protect\citeauthoryear{{Hale} \textit{et~al.}}{1919}]{Hale19}
\begin{barticle}
\bauthor{\bsnm{{Hale}}, \binits{G.E.}},
\bauthor{\bsnm{{Ellerman}}, \binits{F.}},
\bauthor{\bsnm{{Nicholson}}, \binits{S.B.}},
\bauthor{\bsnm{{Joy}}, \binits{A.H.}}:
\byear{1919},
\batitle{{The Magnetic Polarity of Sun-Spots}}.
\bjtitle{\apj}
\bvolume{49},
\bfpage{153}.
\doiurl{https://doi.org/10.1086/142452}.
\adsurl{1919ApJ....49..153H}.
\end{barticle}
\endbibitem

\bibitem[\protect\citeauthoryear{{Hazra}, {Choudhuri}, and
  {Miesch}}{2017}]{Hazra17}
\begin{barticle}
\bauthor{\bsnm{{Hazra}}, \binits{G.}},
\bauthor{\bsnm{{Choudhuri}}, \binits{A.R.}},
\bauthor{\bsnm{{Miesch}}, \binits{M.S.}}:
\byear{2017},
\batitle{{A Theoretical Study of the Build-up of the Sun{\textquoteright}s
  Polar Magnetic Field by using a 3D Kinematic Dynamo Model}}.
\bjtitle{\apj}
\bvolume{835},
\bfpage{39}.
\doiurl{https://doi.org/10.3847/1538-4357/835/1/39}.
\adsurl{2017ApJ...835...39H}.
\end{barticle}
\endbibitem

\bibitem[\protect\citeauthoryear{{Howard}}{1989}]{Howard89}
\begin{barticle}
\bauthor{\bsnm{{Howard}}, \binits{R.F.}}:
\byear{1989},
\batitle{{The Magnetic Fields of Active Regions - Part One}}.
\bjtitle{\solphys}
\bvolume{123},
\bfpage{271}.
\doiurl{https://doi.org/10.1007/BF00149106}.
\adsurl{1989SoPh..123..271H}.
\end{barticle}
\endbibitem

\bibitem[\protect\citeauthoryear{{Jiang}, {Cameron}, and
  {Sch{\"u}ssler}}{2015}]{Jiang15}
\begin{barticle}
\bauthor{\bsnm{{Jiang}}, \binits{J.}},
\bauthor{\bsnm{{Cameron}}, \binits{R.H.}},
\bauthor{\bsnm{{Sch{\"u}ssler}}, \binits{M.}}:
\byear{2015},
\batitle{{The Cause of the Weak Solar Cycle 24}}.
\bjtitle{\apjl}
\bvolume{808},
\bfpage{L28}.
\doiurl{https://doi.org/10.1088/2041-8205/808/1/L28}.
\adsurl{2015ApJ...808L..28J}.
\end{barticle}
\endbibitem

\bibitem[\protect\citeauthoryear{{Kazantsev}}{1968}]{Kazantsev68}
\begin{barticle}
\bauthor{\bsnm{{Kazantsev}}, \binits{A.P.}}:
\byear{1968},
\batitle{{Enhancement of a Magnetic Field by a Conducting Fluid}}.
\bjtitle{Sov. J. Exp. Theor. Phys.}
\bvolume{26},
\bfpage{1031}.
\adsurl{1968JETP...26.1031K}.
\end{barticle}
\endbibitem

\bibitem[\protect\citeauthoryear{{Khlystova} and
  {Sokoloff}}{2009}]{Khlystova09}
\begin{barticle}
\bauthor{\bsnm{{Khlystova}}, \binits{A.I.}},
\bauthor{\bsnm{{Sokoloff}}, \binits{D.D.}}:
\byear{2009},
\batitle{{Toroidal magnetic field of the Sun from data on Hale-rule-violating
  sunspot groups}}.
\bjtitle{Astron. Rep.}
\bvolume{53},
\bfpage{281}.
\doiurl{https://doi.org/10.1134/S106377290903010X}.
\adsurl{2009ARep...53..281K}.
\end{barticle}
\endbibitem

\bibitem[\protect\citeauthoryear{{Kraichnan} and
  {Nagarajan}}{1967}]{Kraichnan67}
\begin{barticle}
\bauthor{\bsnm{{Kraichnan}}, \binits{R.H.}},
\bauthor{\bsnm{{Nagarajan}}, \binits{S.}}:
\byear{1967},
\batitle{{Growth of Turbulent Magnetic Fields}}.
\bjtitle{Phys. Fl.}
\bvolume{10},
\bfpage{859}.
\doiurl{https://doi.org/10.1063/1.1762201}.
\adsurl{1967PhFl...10..859K}.
\end{barticle}
\endbibitem

\bibitem[\protect\citeauthoryear{{Krause} and {R{\"a}dler}}{1980}]{Krause80}
\begin{bbook}
\bauthor{\bsnm{{Krause}}, \binits{F.}},
\bauthor{\bsnm{{R{\"a}dler}}, \binits{K.-H.}}:
\byear{1980},
\bbtitle{{Mean-field magnetohydrodynamics and dynamo theory}}.
\adsurl{1980opp..bookR....K}.
\end{bbook}
\endbibitem

\bibitem[\protect\citeauthoryear{{K{\"u}nzel}}{1965}]{Kuenzel65}
\begin{barticle}
\bauthor{\bsnm{{K{\"u}nzel}}, \binits{H.}}:
\byear{1965},
\batitle{{Zur Klassifikation von Sonnenfleckengruppen}}.
\bjtitle{Astron. Nachr.}
\bvolume{288},
\bfpage{177}.
\adsurl{1965AN....288..177K}.
\end{barticle}
\endbibitem

\bibitem[\protect\citeauthoryear{{Lefevre} and {Clette}}{2014}]{Lefevre14}
\begin{barticle}
\bauthor{\bsnm{{Lefevre}}, \binits{L.}},
\bauthor{\bsnm{{Clette}}, \binits{F.}}:
\byear{2014},
\batitle{{Survey and Merging of Sunspot Catalogs}}.
\bjtitle{\solphys}
\bvolume{289},
\bfpage{545}.
\doiurl{https://doi.org/10.1007/s11207-012-0184-5}.
\adsurl{2014SoPh..289..545L}.
\end{barticle}
\endbibitem

\bibitem[\protect\citeauthoryear{{Leighton}}{1964}]{Leighton64}
\begin{barticle}
\bauthor{\bsnm{{Leighton}}, \binits{R.B.}}:
\byear{1964},
\batitle{{Transport of Magnetic Fields on the Sun.}}
\bjtitle{\apj}
\bvolume{140},
\bfpage{1547}.
\doiurl{https://doi.org/10.1086/148058}.
\adsurl{1964ApJ...140.1547L}.
\end{barticle}
\endbibitem

\bibitem[\protect\citeauthoryear{{Lemen} \textit{et~al.}}{2012}]{Lemen12}
\begin{barticle}
\bauthor{\bsnm{{Lemen}}, \binits{J.R.}},
\bauthor{\bsnm{{Title}}, \binits{A.M.}},
\bauthor{\bsnm{{Akin}}, \binits{D.J.}},
\bauthor{\bsnm{{Boerner}}, \binits{P.F.}},
\bauthor{\bsnm{{Chou}}, \binits{C.}},
\bauthor{\bsnm{{Drake}}, \binits{J.F.}},
\bauthor{\bsnm{{Duncan}}, \binits{D.W.}},
\bauthor{\bsnm{{Edwards}}, \binits{C.G.}},
\bauthor{\bsnm{{Friedlaender}}, \binits{F.M.}},
\bauthor{\bsnm{{Heyman}}, \binits{G.F.}},
\bauthor{\bsnm{{Hurlburt}}, \binits{N.E.}},
\bauthor{\bsnm{{Katz}}, \binits{N.L.}},
\bauthor{\bsnm{{Kushner}}, \binits{G.D.}},
\bauthor{\bsnm{{Levay}}, \binits{M.}},
\bauthor{\bsnm{{Lindgren}}, \binits{R.W.}},
\bauthor{\bsnm{{Mathur}}, \binits{D.P.}},
\bauthor{\bsnm{{McFeaters}}, \binits{E.L.}},
\bauthor{\bsnm{{Mitchell}}, \binits{S.}},
\bauthor{\bsnm{{Rehse}}, \binits{R.A.}},
\bauthor{\bsnm{{Schrijver}}, \binits{C.J.}},
\bauthor{\bsnm{{Springer}}, \binits{L.A.}},
\bauthor{\bsnm{{Stern}}, \binits{R.A.}},
\bauthor{\bsnm{{Tarbell}}, \binits{T.D.}},
\bauthor{\bsnm{{Wuelser}}, \binits{J.-P.}},
\bauthor{\bsnm{{Wolfson}}, \binits{C.J.}},
\bauthor{\bsnm{{Yanari}}, \binits{C.}},
\bauthor{\bsnm{{Bookbinder}}, \binits{J.A.}},
\bauthor{\bsnm{{Cheimets}}, \binits{P.N.}},
\bauthor{\bsnm{{Caldwell}}, \binits{D.}},
\bauthor{\bsnm{{Deluca}}, \binits{E.E.}},
\bauthor{\bsnm{{Gates}}, \binits{R.}},
\bauthor{\bsnm{{Golub}}, \binits{L.}},
\bauthor{\bsnm{{Park}}, \binits{S.}},
\bauthor{\bsnm{{Podgorski}}, \binits{W.A.}},
\bauthor{\bsnm{{Bush}}, \binits{R.I.}},
\bauthor{\bsnm{{Scherrer}}, \binits{P.H.}},
\bauthor{\bsnm{{Gummin}}, \binits{M.A.}},
\bauthor{\bsnm{{Smith}}, \binits{P.}},
\bauthor{\bsnm{{Auker}}, \binits{G.}},
\bauthor{\bsnm{{Jerram}}, \binits{P.}},
\bauthor{\bsnm{{Pool}}, \binits{P.}},
\bauthor{\bsnm{{Soufli}}, \binits{R.}},
\bauthor{\bsnm{{Windt}}, \binits{D.L.}},
\bauthor{\bsnm{{Beardsley}}, \binits{S.}},
\bauthor{\bsnm{{Clapp}}, \binits{M.}},
\bauthor{\bsnm{{Lang}}, \binits{J.}},
\bauthor{\bsnm{{Waltham}}, \binits{N.}}:
\byear{2012},
\batitle{{The Atmospheric Imaging Assembly (AIA) on the Solar Dynamics
  Observatory (SDO)}}.
\bjtitle{\solphys}
\bvolume{275},
\bfpage{17}.
\doiurl{https://doi.org/10.1007/s11207-011-9776-8}.
\adsurl{2012SoPh..275...17L}.
\end{barticle}
\endbibitem

\bibitem[\protect\citeauthoryear{{Li}}{2018}]{Li18}
\begin{barticle}
\bauthor{\bsnm{{Li}}, \binits{J.}}:
\byear{2018},
\batitle{{A Systematic Study of Hale and Anti-Hale Sunspot Physical
  Parameters}}.
\bjtitle{\apj}
\bvolume{867},
\bfpage{89}.
\doiurl{https://doi.org/10.3847/1538-4357/aae31a}.
\adsurl{2018ApJ...867...89L}.
\end{barticle}
\endbibitem

\bibitem[\protect\citeauthoryear{{Li} and {Ulrich}}{2012}]{Li12}
\begin{barticle}
\bauthor{\bsnm{{Li}}, \binits{J.}},
\bauthor{\bsnm{{Ulrich}}, \binits{R.K.}}:
\byear{2012},
\batitle{{Long-term Measurements of Sunspot Magnetic Tilt Angles}}.
\bjtitle{\apj}
\bvolume{758},
\bfpage{115}.
\doiurl{https://doi.org/10.1088/0004-637X/758/2/115}.
\adsurl{2012ApJ...758..115L}.
\end{barticle}
\endbibitem

\bibitem[\protect\citeauthoryear{{L{\'o}pez Fuentes}
  \textit{et~al.}}{2000}]{LopezFuentes00}
\begin{barticle}
\bauthor{\bsnm{{L{\'o}pez Fuentes}}, \binits{M.C.}},
\bauthor{\bsnm{{Demoulin}}, \binits{P.}},
\bauthor{\bsnm{{Mandrini}}, \binits{C.H.}},
\bauthor{\bsnm{{van Driel-Gesztelyi}}, \binits{L.}}:
\byear{2000},
\batitle{{The Counterkink Rotation of a Non-Hale Active Region}}.
\bjtitle{\apj}
\bvolume{544},
\bfpage{540}.
\doiurl{https://doi.org/10.1086/317180}.
\adsurl{2000ApJ...544..540L}.
\end{barticle}
\endbibitem

\bibitem[\protect\citeauthoryear{{L{\'o}pez Fuentes}
  \textit{et~al.}}{2003}]{LopezFuentes03}
\begin{barticle}
\bauthor{\bsnm{{L{\'o}pez Fuentes}}, \binits{M.C.}},
\bauthor{\bsnm{{D{\'e}moulin}}, \binits{P.}},
\bauthor{\bsnm{{Mandrini}}, \binits{C.H.}},
\bauthor{\bsnm{{Pevtsov}}, \binits{A.A.}},
\bauthor{\bsnm{{van Driel-Gesztelyi}}, \binits{L.}}:
\byear{2003},
\batitle{{Magnetic twist and writhe of active regions. On the origin of
  deformed flux tubes}}.
\bjtitle{\aap}
\bvolume{397},
\bfpage{305}.
\doiurl{https://doi.org/10.1051/0004-6361:20021487}.
\adsurl{2003A&A...397..305L}.
\end{barticle}
\endbibitem

\bibitem[\protect\citeauthoryear{{McClintock}, {Norton}, and
  {Li}}{2014}]{McClintock14}
\begin{barticle}
\bauthor{\bsnm{{McClintock}}, \binits{B.H.}},
\bauthor{\bsnm{{Norton}}, \binits{A.A.}},
\bauthor{\bsnm{{Li}}, \binits{J.}}:
\byear{2014},
\batitle{{Re-examining Sunspot Tilt Angle to Include Anti-Hale Statistics}}.
\bjtitle{\apj}
\bvolume{797},
\bfpage{130}.
\doiurl{https://doi.org/10.1088/0004-637X/797/2/130}.
\adsurl{2014ApJ...797..130M}.
\end{barticle}
\endbibitem

\bibitem[\protect\citeauthoryear{{Moffatt}}{1978}]{Moffatt78}
\begin{bbook}
\bauthor{\bsnm{{Moffatt}}, \binits{H.K.}}:
\byear{1978},
\bbtitle{{Magnetic Field Generation in Electrically Conducting Fluids}},
\bpublisher{Cambridge Univ. Press},
\blocation{Cambridge, UK}.
\adsurl{1978mfge.book.....M}.
\end{bbook}
\endbibitem

\bibitem[\protect\citeauthoryear{{Mordvinov} and
  {Kitchatinov}}{2019}]{Mordvinov19}
\begin{barticle}
\bauthor{\bsnm{{Mordvinov}}, \binits{A.V.}},
\bauthor{\bsnm{{Kitchatinov}}, \binits{L.L.}}:
\byear{2019},
\batitle{{Evolution of the Sun's Polar Fields and the Poleward Transport of
  Remnant Magnetic Flux}}.
\bjtitle{\solphys}
\bvolume{294},
\bfpage{21}.
\doiurl{https://doi.org/10.1007/s11207-019-1410-1}.
\adsurl{2019SoPh..294...21M}.
\end{barticle}
\endbibitem

\bibitem[\protect\citeauthoryear{{Moses} \textit{et~al.}}{1997}]{Moses97}
\begin{barticle}
\bauthor{\bsnm{{Moses}}, \binits{D.}},
\bauthor{\bsnm{{Clette}}, \binits{F.}},
\bauthor{\bsnm{{Delaboudini{\`e}re}}, \binits{J.-P.}},
\bauthor{\bsnm{{Artzner}}, \binits{G.E.}},
\bauthor{\bsnm{{Bougnet}}, \binits{M.}},
\bauthor{\bsnm{{Brunaud}}, \binits{J.}},
\bauthor{\bsnm{{Carabetian}}, \binits{C.}},
\bauthor{\bsnm{{Gabriel}}, \binits{A.H.}},
\bauthor{\bsnm{{Hochedez}}, \binits{J.F.}},
\bauthor{\bsnm{{Millier}}, \binits{F.}},
\bauthor{\bsnm{{Song}}, \binits{X.Y.}},
\bauthor{\bsnm{{Au}}, \binits{B.}},
\bauthor{\bsnm{{Dere}}, \binits{K.P.}},
\bauthor{\bsnm{{Howard}}, \binits{R.A.}},
\bauthor{\bsnm{{Kreplin}}, \binits{R.}},
\bauthor{\bsnm{{Michels}}, \binits{D.J.}},
\bauthor{\bsnm{{Defise}}, \binits{J.M.}},
\bauthor{\bsnm{{Jamar}}, \binits{C.}},
\bauthor{\bsnm{{Rochus}}, \binits{P.}},
\bauthor{\bsnm{{Chauvineau}}, \binits{J.P.}},
\bauthor{\bsnm{{Marioge}}, \binits{J.P.}},
\bauthor{\bsnm{{Catura}}, \binits{R.C.}},
\bauthor{\bsnm{{Lemen}}, \binits{J.R.}},
\bauthor{\bsnm{{Shing}}, \binits{L.}},
\bauthor{\bsnm{{Stern}}, \binits{R.A.}},
\bauthor{\bsnm{{Gurman}}, \binits{J.B.}},
\bauthor{\bsnm{{Neupert}}, \binits{W.M.}},
\bauthor{\bsnm{{Newmark}}, \binits{J.}},
\bauthor{\bsnm{{Thompson}}, \binits{B.}},
\bauthor{\bsnm{{Maucherat}}, \binits{A.}},
\bauthor{\bsnm{{Portier-Fozzani}}, \binits{F.}},
\bauthor{\bsnm{{Berghmans}}, \binits{D.}},
\bauthor{\bsnm{{Cugnon}}, \binits{P.}},
\bauthor{\bsnm{{van Dessel}}, \binits{E.L.}},
\bauthor{\bsnm{{Gabryl}}, \binits{J.R.}}:
\byear{1997},
\batitle{{EIT Observations of the Extreme Ultraviolet Sun}}.
\bjtitle{\solphys}
\bvolume{175},
\bfpage{571}.
\doiurl{https://doi.org/10.1023/A:1004902913117}.
\adsurl{1997SoPh..175..571M}.
\end{barticle}
\endbibitem

\bibitem[\protect\citeauthoryear{{Parker}}{1955}]{Parker55}
\begin{barticle}
\bauthor{\bsnm{{Parker}}, \binits{E.N.}}:
\byear{1955},
\batitle{{Hydromagnetic Dynamo Models.}}
\bjtitle{\apj}
\bvolume{122},
\bfpage{293}.
\doiurl{https://doi.org/10.1086/146087}.
\adsurl{1955ApJ...122..293P}.
\end{barticle}
\endbibitem

\bibitem[\protect\citeauthoryear{{R{\"a}dler}, {Kleeorin}, and
  {Rogachevskii}}{2003}]{Radler03}
\begin{barticle}
\bauthor{\bsnm{{R{\"a}dler}}, \binits{K.-H.}},
\bauthor{\bsnm{{Kleeorin}}, \binits{N.}},
\bauthor{\bsnm{{Rogachevskii}}, \binits{I.}}:
\byear{2003},
\batitle{{The Mean Electromotive Force for MHD Turbulence: The Case of a Weak
  Mean Magnetic Field and Slow Rotation}}.
\bjtitle{Geophys. Astrophys. Fluid Dyn.}
\bvolume{97},
\bfpage{249}.
\doiurl{https://doi.org/10.1080/0309192031000151212}.
\adsurl{2003GApFD..97..249R}.
\end{barticle}
\endbibitem

\bibitem[\protect\citeauthoryear{{Richardson}}{1948}]{Richardson48}
\begin{barticle}
\bauthor{\bsnm{{Richardson}}, \binits{R.S.}}:
\byear{1948},
\batitle{{Sunspot Groups of Irregular Magnetic Polarity.}}
\bjtitle{\apj}
\bvolume{107},
\bfpage{78}.
\doiurl{https://doi.org/10.1086/144988}.
\adsurl{1948ApJ...107...78R}.
\end{barticle}
\endbibitem

\bibitem[\protect\citeauthoryear{{Ryutova}}{2018}]{Ryutova18}
\begin{barticle}
\bauthor{\bsnm{{Ryutova}}, \binits{M.}}:
\byear{2018},
\batitle{{Physics of Magnetic Flux Tubes}}.
\bjtitle{Astrophys. Space Sci. Lib.}
\bvolume{455}.
\doiurl{https://doi.org/10.1007/978-3-319-96361-7}.
\adsurl{2018ASSL..455.....R}.
\end{barticle}
\endbibitem

\bibitem[\protect\citeauthoryear{{Scherrer} \textit{et~al.}}{1995}]{Scherrer95}
\begin{barticle}
\bauthor{\bsnm{{Scherrer}}, \binits{P.H.}},
\bauthor{\bsnm{{Bogart}}, \binits{R.S.}},
\bauthor{\bsnm{{Bush}}, \binits{R.I.}},
\bauthor{\bsnm{{Hoeksema}}, \binits{J.T.}},
\bauthor{\bsnm{{Kosovichev}}, \binits{A.G.}},
\bauthor{\bsnm{{Schou}}, \binits{J.}},
\bauthor{\bsnm{{Rosenberg}}, \binits{W.}},
\bauthor{\bsnm{{Springer}}, \binits{L.}},
\bauthor{\bsnm{{Tarbell}}, \binits{T.D.}},
\bauthor{\bsnm{{Title}}, \binits{A.}},
\bauthor{\bsnm{{Wolfson}}, \binits{C.J.}},
\bauthor{\bsnm{{Zayer}}, \binits{I.}},
\bauthor{\bsnm{{MDI Engineering Team}}}:
\byear{1995},
\batitle{{The Solar Oscillations Investigation - Michelson Doppler Imager}}.
\bjtitle{\solphys}
\bvolume{162},
\bfpage{129}.
\doiurl{https://doi.org/10.1007/BF00733429}.
\adsurl{1995SoPh..162..129S}.
\end{barticle}
\endbibitem

\bibitem[\protect\citeauthoryear{{Scherrer} \textit{et~al.}}{2012}]{Scherrer12}
\begin{barticle}
\bauthor{\bsnm{{Scherrer}}, \binits{P.H.}},
\bauthor{\bsnm{{Schou}}, \binits{J.}},
\bauthor{\bsnm{{Bush}}, \binits{R.I.}},
\bauthor{\bsnm{{Kosovichev}}, \binits{A.G.}},
\bauthor{\bsnm{{Bogart}}, \binits{R.S.}},
\bauthor{\bsnm{{Hoeksema}}, \binits{J.T.}},
\bauthor{\bsnm{{Liu}}, \binits{Y.}},
\bauthor{\bsnm{{Duvall}}, \binits{T.L.}},
\bauthor{\bsnm{{Zhao}}, \binits{J.}},
\bauthor{\bsnm{{Title}}, \binits{A.M.}},
\bauthor{\bsnm{{Schrijver}}, \binits{C.J.}},
\bauthor{\bsnm{{Tarbell}}, \binits{T.D.}},
\bauthor{\bsnm{{Tomczyk}}, \binits{S.}}:
\byear{2012},
\batitle{{The Helioseismic and Magnetic Imager (HMI) Investigation for the
  Solar Dynamics Observatory (SDO)}}.
\bjtitle{\solphys}
\bvolume{275},
\bfpage{207}.
\doiurl{https://doi.org/10.1007/s11207-011-9834-2}.
\adsurl{2012SoPh..275..207S}.
\end{barticle}
\endbibitem

\bibitem[\protect\citeauthoryear{{Schunker} \textit{et~al.}}{2020}]{Schunker20}
\begin{barticle}
\bauthor{\bsnm{{Schunker}}, \binits{H.}},
\bauthor{\bsnm{{Baumgartner}}, \binits{C.}},
\bauthor{\bsnm{{Birch}}, \binits{A.C.}},
\bauthor{\bsnm{{Cameron}}, \binits{R.H.}},
\bauthor{\bsnm{{Braun}}, \binits{D.C.}},
\bauthor{\bsnm{{Gizon}}, \binits{L.}}:
\byear{2020},
\batitle{{Average motion of emerging solar active region polarities. II. Joy's
  law}}.
\bjtitle{\aap}
\bvolume{640},
\bfpage{A116}.
\doiurl{https://doi.org/10.1051/0004-6361/201937322}.
\adsurl{2020A&A...640A.116S}.
\end{barticle}
\endbibitem

\bibitem[\protect\citeauthoryear{{Smith} and {Howard}}{1968}]{Smith68}
\begin{bchapter}
\bauthor{\bsnm{{Smith}}, \binits{S.F.}},
\bauthor{\bsnm{{Howard}}, \binits{R.}}:
\byear{1968},
\bctitle{{Magnetic Classification of Active Regions}}.
In: \beditor{\bsnm{{Kiepenheuer}}, \binits{K.O.}} (ed.)
\bbtitle{Structure and Development of Solar Active Regions},
\bsertitle{IAU Symp.}
\bseriesno{35},
\bfpage{33}.
\adsurl{1968IAUS...35...33S}.
\end{bchapter}
\endbibitem

\bibitem[\protect\citeauthoryear{{Sokoloff} and {Khlystova}}{2010}]{Sokoloff10}
\begin{barticle}
\bauthor{\bsnm{{Sokoloff}}, \binits{D.}},
\bauthor{\bsnm{{Khlystova}}, \binits{A.I.}}:
\byear{2010},
\batitle{{The solar dynamo in the light of the distribution of various sunspot
  magnetic classes over butterfly diagram}}.
\bjtitle{Astron. Nachr.}
\bvolume{331},
\bfpage{82}.
\doiurl{https://doi.org/10.1002/asna.200911300}.
\adsurl{2010AN....331...82S}.
\end{barticle}
\endbibitem

\bibitem[\protect\citeauthoryear{{Sokoloff}, {Khlystova}, and
  {Abramenko}}{2015}]{Sokoloff15}
\begin{barticle}
\bauthor{\bsnm{{Sokoloff}}, \binits{D.}},
\bauthor{\bsnm{{Khlystova}}, \binits{A.}},
\bauthor{\bsnm{{Abramenko}}, \binits{V.}}:
\byear{2015},
\batitle{{Solar small-scale dynamo and polarity of sunspot groups}}.
\bjtitle{\mnras}
\bvolume{451},
\bfpage{1522}.
\doiurl{https://doi.org/10.1093/mnras/stv1036}.
\adsurl{2015MNRAS.451.1522S}.
\end{barticle}
\endbibitem

\bibitem[\protect\citeauthoryear{{Stenflo} and {Kosovichev}}{2012}]{Stenflo12}
\begin{barticle}
\bauthor{\bsnm{{Stenflo}}, \binits{J.O.}},
\bauthor{\bsnm{{Kosovichev}}, \binits{A.G.}}:
\byear{2012},
\batitle{{Bipolar Magnetic Regions on the Sun: Global Analysis of the SOHO/MDI
  Data Set}}.
\bjtitle{\apj}
\bvolume{745},
\bfpage{129}.
\doiurl{https://doi.org/10.1088/0004-637X/745/2/129}.
\adsurl{2012ApJ...745..129S}.
\end{barticle}
\endbibitem

\bibitem[\protect\citeauthoryear{{Sur}, {Brandenburg}, and
  {Subramanian}}{2008}]{Sur08}
\begin{barticle}
\bauthor{\bsnm{{Sur}}, \binits{S.}},
\bauthor{\bsnm{{Brandenburg}}, \binits{A.}},
\bauthor{\bsnm{{Subramanian}}, \binits{K.}}:
\byear{2008},
\batitle{{Kinematic {\ensuremath{\alpha}}-effect in isotropic turbulence
  simulations}}.
\bjtitle{\mnras}
\bvolume{385},
\bfpage{L15}.
\doiurl{https://doi.org/10.1111/j.1745-3933.2008.00423.x}.
\adsurl{2008MNRAS.385L..15S}.
\end{barticle}
\endbibitem

\bibitem[\protect\citeauthoryear{{Tlatov}, {Vasil'eva}, and
  {Pevtsov}}{2010}]{Tlatov10}
\begin{barticle}
\bauthor{\bsnm{{Tlatov}}, \binits{A.G.}},
\bauthor{\bsnm{{Vasil'eva}}, \binits{V.V.}},
\bauthor{\bsnm{{Pevtsov}}, \binits{A.A.}}:
\byear{2010},
\batitle{{Distribution of Magnetic Bipoles on the Sun over Three Solar
  Cycles}}.
\bjtitle{\apj}
\bvolume{717},
\bfpage{357}.
\doiurl{https://doi.org/10.1088/0004-637X/717/1/357}.
\adsurl{2010ApJ...717..357T}.
\end{barticle}
\endbibitem

\bibitem[\protect\citeauthoryear{{van Driel-Gesztelyi} and
  {Green}}{2015}]{vanDriel15}
\begin{barticle}
\bauthor{\bsnm{{van Driel-Gesztelyi}}, \binits{L.}},
\bauthor{\bsnm{{Green}}, \binits{L.M.}}:
\byear{2015},
\batitle{{Evolution of Active Regions}}.
\bjtitle{Liv. Rev. Solar Phys.}
\bvolume{12},
\bfpage{1}.
\doiurl{https://doi.org/10.1007/lrsp-2015-1}.
\adsurl{2015LRSP...12....1V}.
\end{barticle}
\endbibitem

\bibitem[\protect\citeauthoryear{{Vitinsky}}{1986}]{Vitinsky86}
\begin{barticle}
\bauthor{\bsnm{{Vitinsky}}, \binits{Y.I.}}:
\byear{1986},
\batitle{{On Sunspot Groups with Irregular Magnetic Polarities}}.
\bjtitle{Byull. Solnechnye Dannye Akad. Nauk SSSR}
\bvolume{9},
\bfpage{86}.
\adsurl{1986BSolD...9...86V}.
\end{barticle}
\endbibitem

\bibitem[\protect\citeauthoryear{{Wang} and {Sheeley}}{1989}]{Wang89}
\begin{barticle}
\bauthor{\bsnm{{Wang}}, \binits{Y.-M.}},
\bauthor{\bsnm{{Sheeley}}, \binits{J.} \bsuffix{N.~R.}}:
\byear{1989},
\batitle{{Average Properties of Bipolar Magnetic Regions during Sunspot
  CYCLE-21}}.
\bjtitle{\solphys}
\bvolume{124},
\bfpage{81}.
\doiurl{https://doi.org/10.1007/BF00146521}.
\adsurl{1989SoPh..124...81W}.
\end{barticle}
\endbibitem

\bibitem[\protect\citeauthoryear{{Wang} \textit{et~al.}}{2015}]{Wang15}
\begin{barticle}
\bauthor{\bsnm{{Wang}}, \binits{Y.-M.}},
\bauthor{\bsnm{{Colaninno}}, \binits{R.C.}},
\bauthor{\bsnm{{Baranyi}}, \binits{T.}},
\bauthor{\bsnm{{Li}}, \binits{J.}}:
\byear{2015},
\batitle{{Active-region Tilt Angles: Magnetic versus White-light Determinations
  of Joy's Law}}.
\bjtitle{\apj}
\bvolume{798},
\bfpage{50}.
\doiurl{https://doi.org/10.1088/0004-637X/798/1/50}.
\adsurl{2015ApJ...798...50W}.
\end{barticle}
\endbibitem

\bibitem[\protect\citeauthoryear{{Yazev}}{2011}]{Yazev11}
\begin{barticle}
\bauthor{\bsnm{{Yazev}}, \binits{S.A.}}:
\byear{2011},
\batitle{{Activity complexes on the sun during the 23rd solar cycle}}.
\bjtitle{Geomag. Aeron.}
\bvolume{51},
\bfpage{879}.
\doiurl{https://doi.org/10.1134/S0016793211070267}.
\adsurl{2011Ge&Ae..51..879Y}.
\end{barticle}
\endbibitem

\bibitem[\protect\citeauthoryear{{Yazev}}{2015}]{Yazev15}
\begin{barticle}
\bauthor{\bsnm{{Yazev}}, \binits{S.A.}}:
\byear{2015},
\batitle{{Activity complexes on the sun in solar cycle 24}}.
\bjtitle{Astron. Rep.}
\bvolume{59},
\bfpage{228}.
\doiurl{https://doi.org/10.1134/S1063772915030075}.
\adsurl{2015ARep...59..228Y}.
\end{barticle}
\endbibitem

\bibitem[\protect\citeauthoryear{{Zeldovich}, {Ruzmaikin}, and
  {Sokoloff}}{1990}]{Zeldovich90}
\begin{bbook}
\bauthor{\bsnm{{Zeldovich}}, \binits{Y.B.}},
\bauthor{\bsnm{{Ruzmaikin}}, \binits{A.A.}},
\bauthor{\bsnm{{Sokoloff}}, \binits{D.D.}}:
\byear{1990},
\bbtitle{{The Almighty Chance}},
\bpublisher{World Scientific},
\blocation{Singapore}.
\doiurl{https://doi.org/10.1142/0862}.
\adsurl{1990alch.book.....Z}.
\end{bbook}
\endbibitem

\bibitem[\protect\citeauthoryear{{Zhukova}}{2018}]{Zhukova18}
\begin{barticle}
\bauthor{\bsnm{{Zhukova}}, \binits{A.}}:
\byear{2018},
\batitle{A catalog of active regions of the 24th solar cycle}.
\bjtitle{Izv. Krym. Astrofiz. Obs.}
\bvolume{114},
\bfpage{74}.
\end{barticle}
\endbibitem

\end{thebibliography}

\end{article} 

\end{document}